\documentclass[12pt,a4paper,epsfig]{article}
\usepackage{amsfonts}
\usepackage{epsfig}

\title{One-loop corrections to the mass \\ of self-dual semi-local planar \\topological solitons}

\author{A. Alonso Izquierdo$^{(a)}$,
W. Garcia Fuertes$^{(b)}$ \\ M. de la Torre Mayado$^{(c)}$, J.
Mateos Guilarte$^{(d)}$
\\ {\normalsize {\it $^{(a)}$ Departamento de Matematica
Aplicada}, {\it Universidad de Salamanca, SPAIN}}\\{\normalsize {\it
$^{(b)}$ Departamento de Fisica} ,{\it Universidad de Oviedo,
SPAIN}}\\ {\normalsize {\it $^{(c)}$ Departamento de Fisica} ,{\it
Universidad de Salamanca, SPAIN}}\\{\normalsize{\it $^{(d)}$
Departamento de Fisica and IUFFyM}, {\it Universidad de Salamanca,
SPAIN}}}
\date{}

\begin{document}
\maketitle
\begin{abstract}
A formula is derived that allows the computation of one-loop mass
shifts for self-dual semilocal topological solitons. These extended
objects, which in three spatial dimensions are called semi-local
strings, arise in a generalized Abelian Higgs model with a doublet
of complex Higgs fields. Having a mixture of global, $SU(2)$, and
local (gauge), $U(1)$, symmetries, this weird system may seem
bizarre, but it is in fact the bosonic sector of electro-weak theory
when the weak mixing angle is ${\pi\over 2}$. The procedure for
computing the semi-classical mass shifts is based on canonical
quantization and heat kernel/zeta function regularization methods.
\end{abstract}

{\footnotesize PACS: 03.70.+k;11.15.Kc;11.15.Ex

}

\section{Introduction}

The purpose of this paper is to address the computation and analysis
of one-loop shifts to the masses of semi-local planar topological
solitons arising in a natural generalization of the Abelian Higgs
model. Seen in (3+1)-dimensional space-time these solitons become
semi-local strings whereas their masses give the string tensions,
see \cite{VaAc}-\cite{VaAc1}. At the critical point that marks the
phase transition between Type I and Type II superconductivity, the
set of semi-local self-dual topological solitons is an interesting
$4l$-dimensional moduli space, where $l$ is the number of quanta of
the magnetic flux, see \cite{GORS}-\cite{JaTa}. Recently,
superconducting semilocal (non self-dual) strings with very
intriguing properties have been discovered in this model
\cite{Forgacs}.

Computations of one-loop mass corrections will be performed using
the heat kernel/zeta function regularization method. The
high-temperature asymptotic expansion of the heat function, see
\cite{Vassilevich}-\cite{Kirsten}-\cite{Ston}-\cite{Gil}, is a
powerful tool that was applied for the first time to the calculation
of kink mass shifts in \cite{Bor} -the ${\cal N}=1$ SUSY case- and
\cite{Aai1} -the non SUSY case-. One-loop corrections to ${\cal
N}=2$ supersymmetric self-dual Nielsen-Olesen vortices were computed
in a similar approach by Vassilevich and Rebhan-van
Nieuwenhuizen-Wimmer in References \cite{Vass} and \cite{Reb}. In
the second paper, the authors also showed that the central charge of
the SUSY algebra is modified in one-loop order in such a way that
the Bogomolny bound is saturated at the semi-classical level.
Sometime later, we calculated the one-loop mass shift for ${\cal
N}=0$ (non-SUSY) self-dual NO vortices carrying a quantum of
magnetic flux in Reference \cite{AGMT}. Mass shifts have been given
for spherically symmetric self-dual vortices (when several solitons
of a quantum of flux have coinciding centers) in \cite{AGMT1} up to
four magnetic flux quanta. Cruder approximations were also provided
for the mass shift of two separated self-dual NO vortices -each of
them with a quantum of magnetic flux- as a function of the
inter-center distance.

In Reference \cite{Aai3}, we studied the one-loop correction to the
energy of a degenerate manifold of kinks that arise in a very
interesting family of models with two real scalar fields in
(1+1)-dimensions. These field theoretical systems are obtained
through dimensional reduction -plus a reality condition- of an
${\cal N}=1$ supersymmetric Wess-Zumino model with two chiral
super-fields, see \cite{Bazeia}-\cite{Bazeia1}-\cite{Shifman}. A
comparison between the mass shifts of these composite kinks and the
correction to the mass of the ordinary $\lambda\phi_2^4$ kink was
offered in \cite{AGMT2}. In this paper, we address a similar, but
more difficult, situation in (2+1)-dimensions, comparing one-loop
mass corrections of the topological solitons that arise in two
planar Abelian gauge systems; one with two complex scalar fields and
the other with a single complex scalar field. The methodology used
to accomplish this task is explained in detail in Reference
\cite{AGGMMT}, where a complete list of References can be found.

Our paper is organized as follows: in Section \S 2 we describe the
model and develop perturbation theory around one of the vacua. A
one-loop renormalization is also performed. Section \S 3 is devoted
to summarizing the structure of the moduli space of self-dual
topological solitons. As a novelty, we also apply a variation of the
de Vega-Shaposnik method \cite{VeSh} to find numerical solutions for
spherically symmetric topological solitons. In Section \S 4, we
explain how to obtain one-loop mass shifts in terms of generalized
zeta functions of the second-order differential operators ruling the
small fluctuations of the bosonic and ghost fields, and in \S 5 the
high-temperature expansion of the heat kernel is used to give the
final formula for one-loop mass shifts of semi-local self-dual
topological solitons after application of Mellin's transforms. We
present our results in Section \S 6 by means of Mathematica
calculations of the coefficients of the asymptotic series giving the
heat functions. Finally, in the Appendix we offer several Tables
where these coefficients are shown.

\section{The planar semilocal Abelian Higgs model}
\subsection*{The Model}
The semi-local Abelian Higgs model \cite{VaAc} describes the minimal
coupling between an $U(1)$-gauge field $A_\mu$ and a doublet $\Phi$
of complex scalar fields in a phase where the Higgs mechanism takes
place. The term semilocal refers to the fact that while the global
symmetry of this system is $SU(2)\times U(1)$ only the $U(1)$ factor
is gauged. Defining non-dimensional space-time variables, $x^\mu
\rightarrow \frac{1}{ev} x^\mu$, and fields, $\Phi \rightarrow
v\Phi$, $A_\mu \rightarrow v A_\mu$, from the vacuum expectation
value of the Higgs field $v$ and the $U(1)$-gauge coupling constant,
$e$, the action for the semilocal Abelian Higgs model in
(2+1)-dimensions reads:
\[S= \frac{v}{e}\int d^3 x \left[
-\frac{1}{4} F_{\mu \nu} F^{\mu \nu}+\frac{1}{2} (D_\mu
\Phi)^\dagger D^\mu \Phi - \frac{\kappa^2}{8} (\Phi^\dagger
\Phi(x^\mu)-1)^2 \right] \qquad ,
\]
where the covariant derivative is defined as
$D_\mu\Phi=(\frac{\partial}{\partial x^\mu}-iA_\mu)\Phi$. The
physical spectrum differs from that of the standard AHM model in
that, along with the massive vector boson and Higgs scalar, there is
a complex Goldstone field. The parameter
$\kappa^2=\frac{\lambda}{e^2}$ measures the ratio between the square
of the masses of the Higgs, $M^2=\lambda v^2$, and vector particles,
$m^2=e^2v^2$. Here, $\lambda$ is the Higgs field self-coupling. We
choose a system of units where $c=1$, but $\hbar$ has dimensions of
length $\times$ mass.

\subsection*{Feynman Rules in the Feynman-'t Hooft R gauge}

In order to obtain the Feynman rules, we expand the action around a
classical vacuum. We shall profit from the $SU(2)$ global invariance
and we choose a vacuum $\Phi^V$ with real upper and vanishing lower
components. The shift of the fields
\[
\Phi(x^\mu)=\left(\begin{array}{c} 1+H(x^\mu)+iG(x^\mu) \\
\sqrt{2}\varphi(x^\mu)\end{array}\right)
\]
corresponds to the identification of $H(x^\mu)$, $G(x^\mu)$, and
$\varphi(x^\mu)$, respectively as  Higgs, real, and complex
Goldstone fields. The choice of the Feynman-'t Hooft R-gauge
\[
R(A_\mu , G)=\partial_\mu A^\mu + G
\]
requires a Faddeev-Popov determinant to restore unitarity, which
amounts to introducing a complex ghost field $\chi$. All this
together allows us to write the action in the form
\begin{eqnarray*}
S&=&{v\over e}\int \, d^3x \, \left[ -\frac{1}{2} A_\mu
[-g^{\mu\nu}(\Box +1)]A_\nu +\partial_\mu\chi^*\partial^\mu \chi- \chi^*\chi\right.\\
&+&\frac{1}{2}\partial_\mu G\partial^\mu G-\frac{1}{2} G^2+
\frac{1}{2}\partial_\mu H\partial^\mu H-\frac{\kappa^2}{2}  H^2+\partial_\mu\varphi^* \partial^\mu \varphi\\
&-& {\kappa^2\over 2}H (H^2+G^2)+ A_\mu (\partial^\mu H
G-\partial^\mu G H)+H (A_\mu A^\mu -\chi^* \chi) +i A_\mu(\varphi^*
\partial^\mu \varphi-\varphi \partial^\mu \varphi^*)\\&+& \left.
A_\mu A^\mu |\varphi|^2 -\frac{\kappa^2}{8} (H^2+G^2)^2
+\frac{1}{2}(G^2+H^2) A_\mu A^\mu
-\frac{\kappa^2}{2}|\varphi|^2(|\varphi|^2+H^2+G^2+2H)\right]
\end{eqnarray*}
in order to deduce the Feynman rules for the propagators and
vertices shown in Tables 1 and 2. The propagator of the complex
Goldstone boson plus two trivalent and four tetravalent vertices
accounting for the interactions of Goldstone-anti-Goldstone pairs
must be added to the Feynman rules of the Abelian Higgs model in the
Feynman-'t Hooft gauge. Nevertheless, in order to guarantee that all
physical quantities will be ultraviolet finite there is still a
remaining piece to be added: the action for the counter-terms. We
shall compute this up to one-loop order in the sequel.

{\footnotesize\begin{table}[h]
\begin{center}
\caption{Propagators}
\begin{tabular}{lccc} \\ \hline
\textit{Particle} & \textit{Field} & \textit{Propagator} & \textit{Diagram} \\
\hline \\ Higgs & $H(x)$ & $\displaystyle\frac{i e \hbar
}{v(k^2-\kappa^2+i\varepsilon)}$ &
\parbox{1.8cm}{\epsfig{file=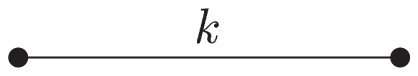,width=1.8cm}} \\[0.5cm]
Real Goldstone & $G(x)$ & $\displaystyle\frac{ie
\hbar}{v(k^2-1+i\varepsilon)}$ &
\parbox{1.8cm}{\epsfig{file=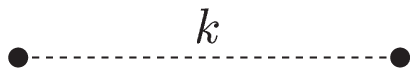,width=1.8cm}} \\[0.5cm]
Complex Goldstone & $\varphi(x)$ & $\displaystyle\frac{ie
\hbar}{v(k^2+i\varepsilon)}$ & \hspace{0.05cm}
\parbox{1.8cm}{\epsfig{file=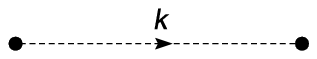,width=1.65cm}} \\[0.5cm]
Ghost & $\chi(x)$ & $\displaystyle\frac{ie
\hbar}{v(k^2-1+i\varepsilon)}$ &
\parbox{1.8cm}{\epsfig{file=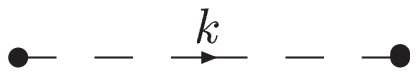,width=1.8cm}} \\[0.5cm]
Vector Boson & $A_\mu(x)$  &  {\Large $\frac{-ie \hbar g^{\mu\nu}
}{v(k^2-1+i\varepsilon)}$} &
\parbox{1.8cm}{\epsfig{file=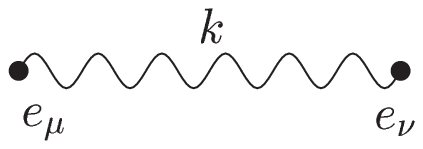,width=2cm}}
\\[0.5cm] \hline
\end{tabular}
\end{center}
\end{table}}
{\footnotesize\begin{table}[hbt]
\begin{center}
\caption{Third- and fourth-order vertices }
\begin{tabular}{llllllll} \\ \hline
\textit{Vertex} & \textit{Weight} & \textit{Vertex} &
\textit{Weight}
& \textit{Vertex} & \textit{Weight} & \textit{Vertex} & \textit{Weight} \\
\hline \\
\parbox{1.8cm}{\epsfig{file=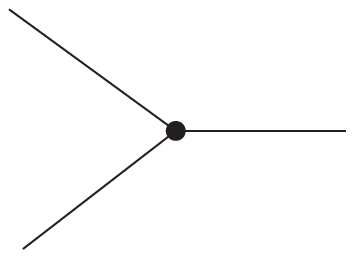,width=1.8cm}} &
$-3i\kappa^2\frac{v}{\hbar e} $  &
\parbox{1.8cm}{\epsfig{file=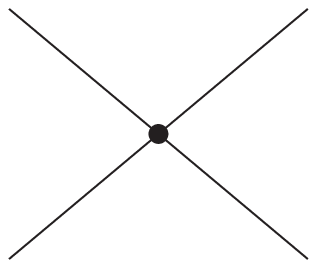,width=1.5cm}} &
$ -3i\kappa^2\frac{v}{\hbar e}$ &
\parbox{1.8cm}{\epsfig{file=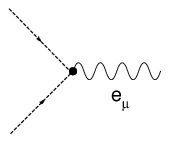,width=1.45cm}} &
$ i(k^\mu+q^\mu)\frac{v}{\hbar e} $  &
\parbox{1.8cm}{\epsfig{file=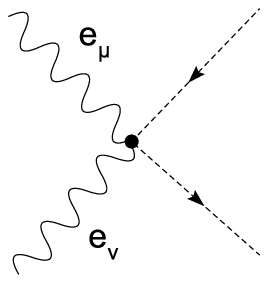,width=1.2cm}} &
$ 2 i g^{\mu\nu}\frac{v}{\hbar e}$ \\[0.5cm]
\parbox{1.8cm}{\epsfig{file=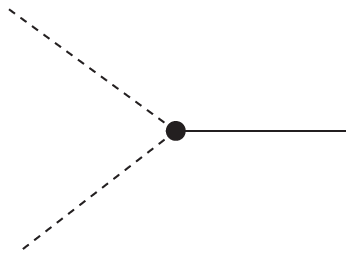,width=1.8cm}} &
$ -i\kappa^2\frac{v}{\hbar e}$  &
\parbox{1.8cm}{\epsfig{file=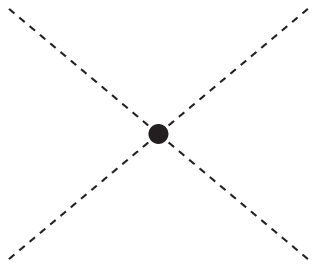,width=1.5cm}} &
$-3i\kappa^2\frac{v}{\hbar e}$ &
\parbox{1.8cm}{\epsfig{file=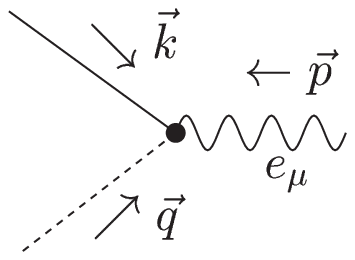,width=1.7cm}} & $
(k^\mu-q^\mu)\frac{v}{\hbar e}$  &
\parbox{1.8cm}{\epsfig{file=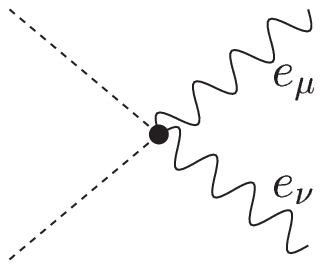,width=1.6cm}} &
$ 2 i \frac{v}{\hbar e} g^{\mu \nu}$ \\[0.5cm]
 \parbox{1.8cm}{\epsfig{file=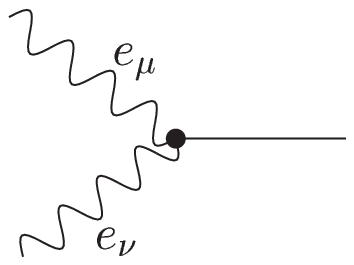,width=1.8cm}} &
$ 2i\frac{v}{\hbar e} g^{\mu \nu}$  &
\parbox{1.8cm}{\epsfig{file=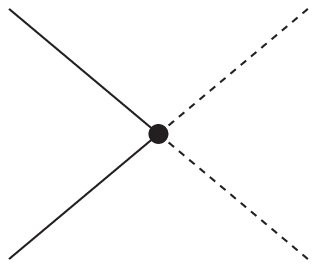,width=1.5cm}} &
$ -i\kappa^2\frac{v}{\hbar e}$ & \hspace{0.1cm}
\parbox{1.8cm}{\epsfig{file=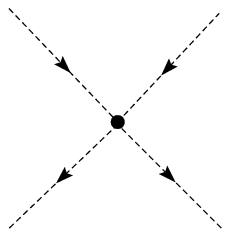,width=1.2cm}} &
$ -2i\kappa^2\frac{v}{\hbar e} $  &
\parbox{1.8cm}{\epsfig{file=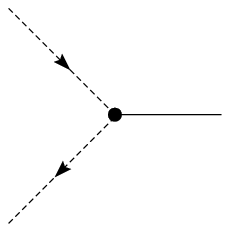,width=1.3cm}} &
$ -i\kappa^2\frac{v}{\hbar e}$
\\[0.5cm]
 \parbox{1.8cm}{\epsfig{file=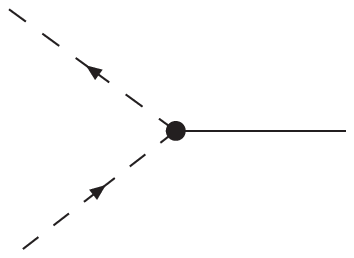,width=1.8cm}} &
$-i\frac{v}{\hbar e}$ &
\parbox{1.8cm}{\epsfig{file=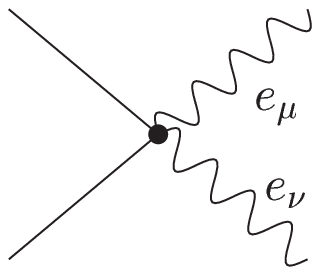,width=1.8cm}} &
$ 2 i\frac{v}{\hbar e}g^{\mu \nu}$ &\hspace{0.1cm}
\parbox{1.8cm}{\epsfig{file=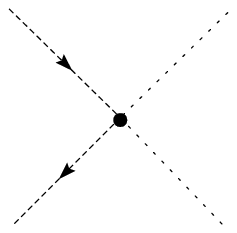,width=1.2cm}} &
$ -i \kappa^2\frac{v}{\hbar e}$ &
\parbox{1.8cm}{\epsfig{file=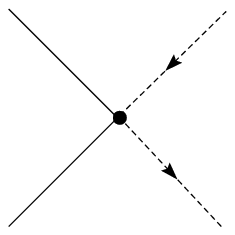,width=1.2cm}} &
$ -i \kappa^2\frac{v}{\hbar e}$  \\[0.5cm]
\\[0.7cm]\hline
\end{tabular}
\end{center}
\end{table}}

\subsection*{One-loop mass renormalization counter-terms}

All ultraviolet divergences come from the integral:
\[
I(c^2)=\int \, \frac{d^3k}{(2\pi)^3} \cdot
\frac{i}{k^2-c^2+i\varepsilon}\qquad .
\]
We also note that even if there are massless particles propagating
in 2+1 dimensions, $I(0)$ is infrared convergent. The
renormalization of the semilocal AHM requires the computation of the
following one-loop graphs:
\begin{enumerate}
\item Higgs boson tadpole:
\vspace{0.4cm}

\centerline{
\parbox{1.5cm}{\epsfig{file=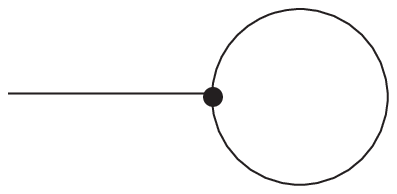,width=1.5cm}} $+$
\parbox{1.5cm}{\epsfig{file=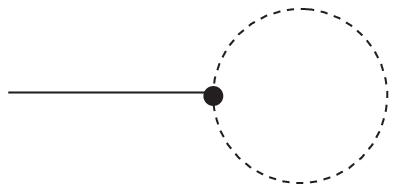,width=1.5cm}} $+$
\parbox{1.5cm}{\epsfig{file=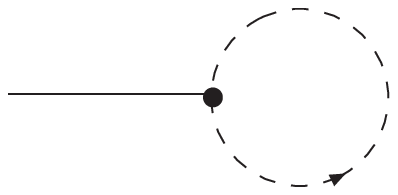,width=1.5cm}} $+$
\parbox{1.5cm}{\epsfig{file=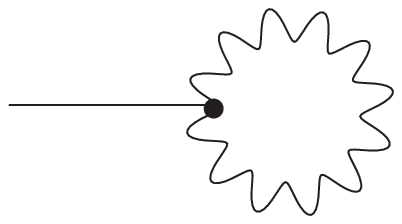,width=1.4cm}} $+$
\parbox{1.5cm}{\epsfig{file=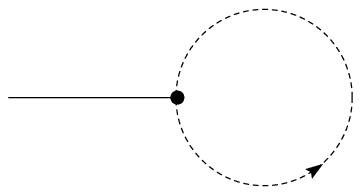,width=1.25cm}}$=$}
\vspace{-0.4cm}
\[
= - 2 i (\kappa^2 + 1)\cdot\, I(1)-i\kappa^2\cdot\, I(0) +
\mbox{finite part}
\]

\item Higgs boson self-energy:
\vspace{0.4cm}

\centerline{
\parbox{1.5cm}{\epsfig{file=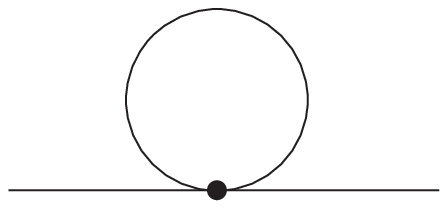,width=1.5cm}} $+$
\parbox{1.5cm}{\epsfig{file=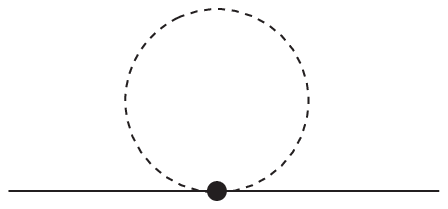,width=1.5cm}} $+$
\parbox{1.5cm}{\epsfig{file=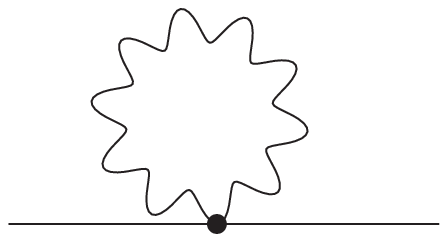,width=1.5cm}} $+$
\parbox{1.5cm}{\epsfig{file=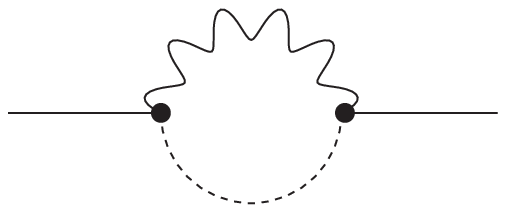,width=1.5cm}}$+$
\parbox{1.5cm}{\epsfig{file=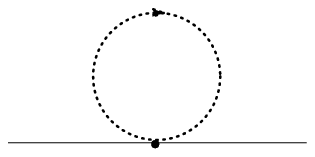,width=1.35cm}} $=$}
\vspace{-0.2cm}
\[
= - 2 i (\kappa^2+ 1)\cdot\, I(1)-i\kappa^2\cdot\, I(0) +
\mbox{finite part}
\]

\item Real Goldstone boson self-energy:
\vspace{0.4cm}

\centerline{
\parbox{1.5cm}{\epsfig{file=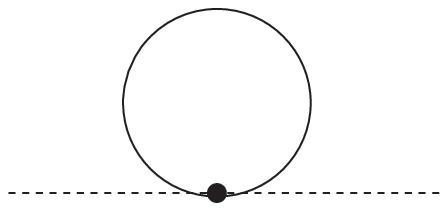,width=1.5cm}} $+$
\parbox{1.5cm}{\epsfig{file=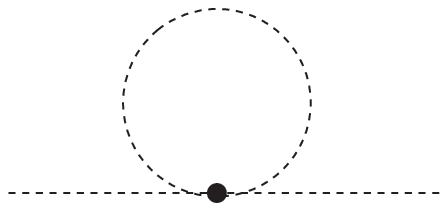,width=1.5cm}} $+$
\parbox{1.5cm}{\epsfig{file=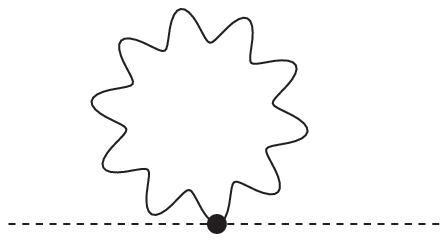,width=1.5cm}} $+$
\parbox{1.5cm}{\epsfig{file=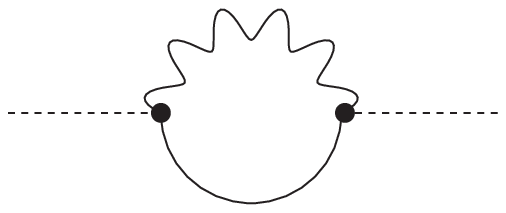,width=1.5cm}}$+$
\parbox{1.5cm}{\epsfig{file=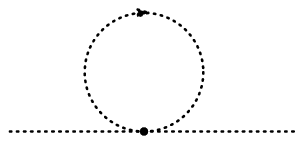,width=1.35cm}}$=$}
\vspace{-0.2cm}
\[
= - 2 i (\kappa^2+ 1)\cdot\, I(1)-i\kappa^2\cdot\, I(0) +
\mbox{finite part}
\]
\item Complex Goldstone boson self-energy:
\vspace{0.4cm}

\centerline{
\parbox{1.5cm}{\epsfig{file=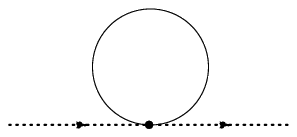,width=1.5cm}} $+$
\parbox{1.5cm}{\epsfig{file=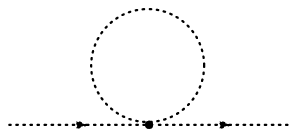,width=1.5cm}} $+$
\parbox{1.5cm}{\epsfig{file=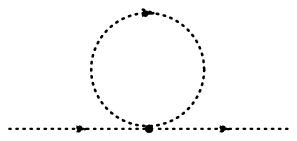,width=1.5cm}} $+$
\parbox{1.5cm}{\epsfig{file=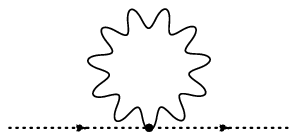,width=1.5cm}}
$+$ \parbox{1.5cm}{\epsfig{file=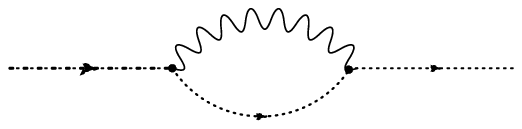,width=2.0cm}}
\qquad $=$} \vspace{-0.2cm}
\[
= - 4i (\kappa^2+ 1)\cdot\, I(1)-2i\kappa^2\cdot\, I(0) +
\mbox{finite part}
\]

\item  Vector boson self-energy: The potentially divergent part is
\vspace{0.4cm}

\centerline{
\parbox{1.5cm}{\epsfig{file=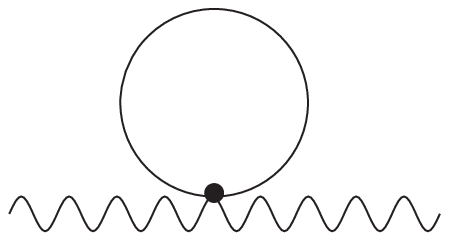,width=1.5cm}} $+$
\parbox{1.5cm}{\epsfig{file=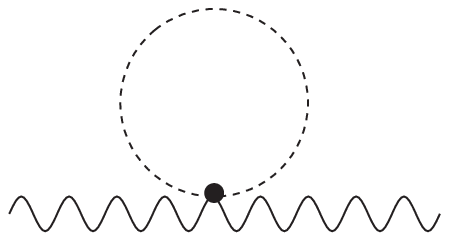,width=1.5cm}} $+$
\parbox{1.6cm}{\epsfig{file=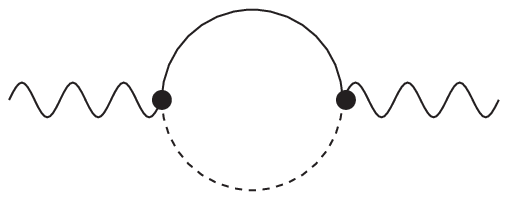,width=1.6cm}} $+$
\parbox{1.5cm}{\epsfig{file=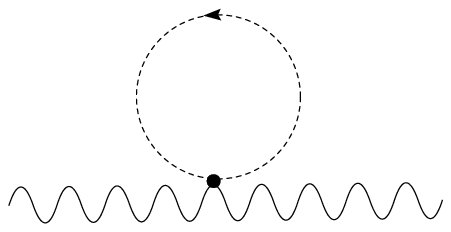,width=1.5cm}}$+$
\parbox{1.7cm}{\epsfig{file=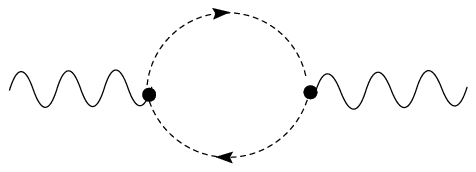,width=1.7cm}} }
\vspace{-0.4cm}
\[
=  2i\cdot\, [I(1)+I(0)]g^{\mu\nu} + \mbox{finite part} \qquad .
\]

\end{enumerate}

In (2+1)-dimensions the graphs above are the only ultraviolet
divergent diagrams for any number of loops in the diagrams, not only
in one-loop order, and the theory is super-renormalizable. Thus, in
a minimal subtraction scheme we get rid off all ultraviolet
divergences arising in the vacuum sector of the model by adding the
counter-terms
\begin{eqnarray*}
{\cal L}_{c.t.}^S &=& {\hbar\over 2}\left[2(\kappa^2+1)\cdot\,
I(1)+\kappa^2\cdot\, I(0)\right]\cdot\,
\left[\Phi^*_1(x^\mu)\Phi_1(x^\mu)+\Phi^*_2(x^\mu)\Phi_2(x^\mu)-1
\right]\\{\cal L}_{c.t.}^A &=&-\hbar [I(1)+I(0)]\cdot\, A_\mu(x^\mu)
A^\mu(x^\mu)
\end{eqnarray*}

\begin{table}[hbt]
\begin{center}
\caption{Counter-term vertices}
\begin{tabular}{llc} \\ \hline
\textit{Diagram} & \hspace{0.3cm} & \textit{Weight} \\
\hline
\parbox{3.5cm}{\epsfig{file=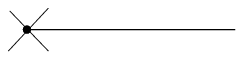,width=3.5cm}} & &
$\displaystyle i[2(\kappa^2+1)I(1)+\kappa^2 I(0)] $
\\
\parbox{3.5cm}{\epsfig{file=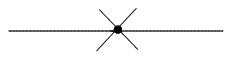,width=3.5cm}} & &
$\displaystyle i[2(\kappa^2+1) I(1)+\kappa^2 I(0)]$ \\
 \parbox{4.3cm}{\epsfig{file=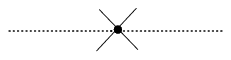,width=3.5cm}} & &
$\displaystyle i[2(\kappa^2+1) I(1)+\kappa^2 I(0)]$
\\
 \parbox{3.5cm}{\epsfig{file=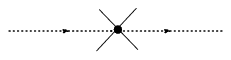,width=3.5cm}} & &
$\displaystyle i[4(\kappa^2+1) I(1)+2\kappa^2 I(0)]$
\\
 \parbox{3.5cm}{\epsfig{file=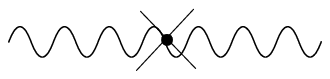,width=3.5cm}} & &
$\displaystyle -2i [I(1)+I(0)]g^{\mu\nu}$ \\[0.5cm] \hline \\[0.25cm]
\end{tabular}
\qquad ,
\end{center}
\end{table}
\noindent which gives rise to the vertices shown in Table 3. In our
renormalization scheme, finite renormalizations have been adjusted
in such a way that, on one hand, the divergence due to the tadpole
graph is exactly canceled when the theory is fine-tuned to the so
called self-dual limit $\kappa^2=1$ and, on the other hand, the
global $SU(2)$ symmetry remains unbroken up to one-loop order.

\section{Semilocal vortices}
\subsection*{Finite energy solutions}
Apart from the ground states built around the $S^3$ classical
homogeneous solutions, the semilocal AHM has room for other quantum
states arising from the extended classical configurations, which are
stable owing to topological reasons. These configurations are the
topological solitons that appear at tree level as non-homogeneous
solutions of the static field equations:
\[
\partial_i F_{ij}=\frac{i}{2} \left( \Phi^\dagger D_j \Phi - (D_j
\Phi)^\dagger\right)\qquad , \qquad  D_i
D_i\Phi=\frac{\kappa^2}{4}\Phi(\Phi^\dagger\Phi-1) \qquad ,
\]
such that their energy,
\[
E= \int d^2 x [\frac{1}{4} F_{ij} F_{ij} + \frac{1}{2} (D_i
\Phi)^\dagger D_i \Phi + \frac{\kappa^2}{8} (\Phi^\dagger \Phi-1)^2
] \nonumber \qquad ,
\]
is finite. The topological character of these solutions is peculiar
in that although the scalar vacuum manifold is $S^3$, and thus
simply connected, the configuration space ${\cal C}$
\[
{\cal C}=\left\{\Phi(\vec{x})\in Maps({\mathbb R}^2,{\mathbb
C}^2),A_i(\vec{x})\in Maps({\mathbb R}^2,T{\mathbb
R}^2)/E(\Phi,A_i)<+\infty \right\}
\]
is the union of ${\mathbb Z}$ disconnected sectors. Finite energy
configurations show the asymptotic behavior
\[
\Phi^{\dagger} \Phi |_{S_\infty^1} =1 \hspace{2cm} , \hspace{2cm}
D_i \Phi |_{S_\infty^1} =(\partial_i \Phi - i A_i
\Phi)|_{S_\infty^1}=0,
\]
where $S_\infty^1$ is the circle that bounds the plane at infinity.
Parametrizing $S^1_\infty$ by the polar angle $\theta={\rm
arctan}\frac{x_2}{x_1}$, and, modulo the global $SU(2)$-symmetry,
choosing $\Phi|_{S_\infty^1}=\Phi^V$ for $\theta=0$, the boundary
conditions on the covariant derivatives provide a map, $S^1_{\infty}
\longrightarrow S^1_1$, between the sphere at the infinity spatial
and the fiber $S_1^1$ at the north pole of the base $S^2$ of the
$S^3$ vacuum Hopf bundle:
\[
\Phi|_{S_\infty^1}=\left(\begin{array}{c}\Phi_1|_{S_\infty^1}\\
\Phi_2|_{S_\infty^1}\end{array}\right)=\left(\begin{array}{c}\phi_1|_{S_\infty^1}+i\phi_2|_{S_\infty^1}\\
\phi_3|_{S_\infty^1}+i\phi_4|_{S_\infty^1}\end{array}\right) =
\left(\begin{array}{c}e^{il \theta}\\0\end{array}\right)\, \,\qquad
, \, \, \qquad l\in{\mathbb Z}\qquad .
\]
Continuous maps between one-dimensional spheres are classified
according to the first homotopy group and, because the temporal
evolution is continuous, $ \Pi_0({\cal C})=\Pi_1(S^1_1)={\mathbb
Z}$, the zero homotopy group of ${\cal C}$ is non-trivial. Thus, $
{\cal C}=\sqcup_{l\in{\mathbb Z}}{\cal C}_l$ is the union of
disconnected sectors characterized by an integer number $l$.

Moreover, the boundary condition for the vector field $A_i
|_{S_\infty^1} =- i \Phi^\dagger\partial_i \Phi|_{S_\infty^1}$ is
tantamount to:
\[
A_i(\vec{x})\left
|_{S^1_\infty}\right.=\phi_1(\vec{x})\frac{\partial\phi_2}{\partial
x_i}(\vec{x})\left
|_{S^1_\infty}\right.-\phi_2(\vec{x})\frac{\partial\phi_1}{\partial
x_i}(\vec{x})\left
|_{S^1_\infty}\right.+\phi_3(\vec{x})\frac{\partial\phi_4}{\partial
x_i}(\vec{x})\left
|_{S^1_\infty}\right.-\phi_4(\vec{x})\frac{\partial\phi_3}{\partial
x_i}(\vec{x})\left |_{S^1_\infty}\right.
\]
The $SU(2)$-orbit of $\Phi_l^V=\left(\begin{array}{c} e^{il\theta}
\\ 0 \end{array}\right)$ is:
\[
G\Phi_l^V=\left(\begin{array}{cc} e^{i\theta_1}{\rm sin}\psi &
e^{i\theta_2}{\rm cos}\psi \\ -e^{-i\theta_2}{\rm cos}\psi &
e^{-i\theta_1}{\rm sin}\psi
\end{array}\right)\left(\begin{array}{c} e^{il\theta} \\ 0
\end{array}\right)=\left(\begin{array}{c} e^{i(\theta_1+l\theta)}{\rm sin}\psi \\
-e^{-i(\theta_2-l\theta)}{\rm cos}\psi  \end{array}\right) \qquad ,
\]
where $\theta_1, \theta_2\in[0,2\pi]$ and $\psi\in[0,{\pi\over 2}]$
are the Hopf coordinates of the $S^3$ sphere. Therefore,
\begin{eqnarray*}
A_i(\vec{x})\left|_{S^1_\infty}\right.&=&l\frac{\partial\theta}{\partial
x_i}\left({\rm cos}^2(\theta_1+l\theta){\rm sin}^2\psi+{\rm
sin}^2(\theta_1+l\theta){\rm sin}^2\psi\right) + \\ &+&
l\frac{\partial\theta}{\partial x_i}\left({\rm
cos}^2(\theta_2-l\theta){\rm cos}^2\psi+{\rm
sin}^2(\theta_2-l\theta){\rm
cos}^2\psi\right)\\&=&l\frac{\partial\theta}{\partial x_i}
\end{eqnarray*}
and the topological (winding) number $l$ has a direct physical
interpretation in terms of the magnetic flux carried by the planar
soliton:
\[
g=\oint_{S^1_\infty}\,
\left(A_1(\vec{x})dx^1+A_2(\vec{x})dx^2\right)=l\oint\,
\left(\frac{\partial\theta}{\partial
x_1}dx^1+\frac{\partial\theta}{\partial
x_2}dx^2\right)=l\int_0^{2\pi}\, d\theta =2\pi l \qquad  .
\]

\subsection*{Self-dual semi-local topological solitons}

We shall restrict ourselves to the critical point between Type I and
Type II superconductivity: $\kappa^2=1$. The energy can be arranged
in a Bogomolny splitting:
\[
E= \int \frac{d^2 x}{2} \left( (D_1 \Phi \pm i D_2 \Phi)^\dagger
(D_1 \Phi \pm i D_2 \Phi) + [ F_{12} \pm {\textstyle\frac{1}{2}}
(\Phi^\dagger \Phi-1) ]^2 \right)+\frac{1}{2}|g|
\]
One immediately realizes that the solutions of the first-order
equations
\begin{equation}
D_1 \Phi \pm i D_2 \Phi=0 \hspace{1.5cm},\hspace{1.5cm} F_{12} \pm
\frac{1}{2} (\Phi^\dagger\Phi-1) =0 \label{eq:for}
\end{equation}
are absolute minima of the energy, and are hence stable, in each
topological sector that has a classical mass proportional to the
magnetic flux. Because the first-order equations can be obtained
from the self-duality equations of Euclidean 4D gauge theory through
dimensional reduction, the vortex solutions of (\ref{eq:for}) are
called self-dual at the limit $\kappa^2=1$.

We follow \cite{GORS} to summarize the properties and existence of
the so-called semi-local self-dual topological solitons: the
solutions of (\ref{eq:for}) for non-negative $l$ (plus sign in the
first-order equations){\footnote{It is trivial to obtain the
solutions for negative $l$ if the solutions for positive $l$ are
known; simply, complex conjugation gives the solution of equation
(\ref{eq:for}) with the other signs.}}. The equation on the left in
(\ref{eq:for}) is tantamount to:
\begin{equation}
\bar{A}(z,\bar{z})=-i\partial_{\bar{z}}\, \log \, \Phi_a(z,\bar{z})
\quad , \quad a=1,2 \qquad \qquad , \qquad
\Phi(z,\bar{z})=\left(\begin{array}{c}\Phi_1(z,\bar{z}) \\
\Phi_2(z,\bar{z})\end{array}\right) \qquad , \label{eq:for1c}
\end{equation}
where the complex notation for coordinates and fields is:
\[
z=x_1+ix_2 \qquad ,\qquad
\partial_z={1\over 2}(\partial_1-i\partial_2) \qquad ,\qquad
A={1\over 2}(A_1-iA_2) \qquad .
\]
From (\ref{eq:for1c}), one sees that:
\[
\partial_{\bar z} \, \log
\left(\frac{\Phi_2(z,\bar{z})}{\Phi_1(z,\bar{z})}\right)=0 \qquad
\Longrightarrow \qquad
\omega(z)=\frac{\Phi_2(z,\bar{z})}{\Phi_1(z,\bar{z})}=\frac{Q_m(z)}{P_l(z)}
\]
where $Q_m$ and $P_l$ are polynomials of respective degree $m$ and
$l$ in $z$, such that $\omega(z)$ is locally analytic. The behavior
of the Higgs field at infinity (up to global $SU(2)$
transformations) compatible with finite energy requires that $m<l$
and $P_l(z)$ be monic:
\[
P_l(z)=z^l+p_{l-1}z^{l-1}+ \cdots +p_1z+p_0 \qquad , \qquad
Q_{l-1}(z)=q_{l-1}z^{l-1}+ \cdots +q_1z+q_0 \qquad .
\]
Therefore, the moduli space of semi-local topological solitons
depends on $2l$ complex ($4l$ real) parameters: $(p_a,q_a), a=0,1,
\cdots , l-1$. Varying the values of $(p_a,q_a)$ one varies the $l$
zeroes of $\Phi_1$ and the $l-1$ zeroes of $\Phi_2$ in such a way
that the locations of the zeroes of the two Higgs fields plus the
scale and orientation of $\Phi_2$ parametrize the SSTS moduli space.
Equation (\ref{eq:for}) on the right becomes
\begin{equation}
\bigtriangleup \, u(z,\bar{z})+1-e^{u(z,\bar{z})}=\bigtriangleup \,
\log \, \left(\left|P_l\right|^2(z)+\left|Q_{l-1}\right|^2(z)\right)
\label{eq:for2c} \qquad ,
\end{equation}
where
\[
u(z,\bar{z})=\Omega(z,\bar{z})+ \log \, (1+|\omega |^2(z)) \qquad ,
\qquad \Omega(z,\bar{z})=\log \, \Phi_1^*\Phi_1(z,\bar{z}) \qquad .
\]
By functional analysis techniques, it is possible to demonstrate the
existence and uniqueness of a solution of (\ref{eq:for2c})
compatible with finite energy boundary conditions \cite{GORS, JaTa}.
Therefore, the finite energy solutions of (\ref{eq:for}) with a
magnetic flux $2\pi l$ take the form
\[
\left(\begin{array}{c} \Phi_1^{(l)}(z,\bar{z}) \\
\Phi_2^{(l)}(z,\bar{z})
\end{array}\right)=\frac{1}{\sqrt{\left|P_l\right|^2(z)+\left|Q_{l-1}\right|^2(z)}}\cdot
\left(\begin{array}{c} P_l(z) \\ Q_{l-1}(z)\end{array}\right)\cdot
e^{{1\over 2}u(z,\bar{z})} \qquad ,
\]
where $u(z,\bar{z})$ is a solution of (\ref{eq:for2c}). In the $l=1$
case we have that:
\[
\left(\begin{array}{c} \Phi_1^{(l=1)}(z,\bar{z}; q_0) \\
\Phi_2^{(l=1)}(z,\bar{z}; q_0)
\end{array}\right)=\frac{1}{\sqrt{\left|z-z_0\right|^2+\left|q_0\right|^2}}\cdot
\left(\begin{array}{c} z-z_0 \\ q_0\end{array}\right)\cdot
e^{{1\over 2}u(|z-z_0|;\,\, q_0)} \qquad .
\]
Setting the parameter $q_0$ to zero, we find the embedded
Nielsen-Olesen vortex centered at $z_0$:
\[
\left(\begin{array}{c} \Phi_1^{(l=1)}(z,\bar{z}; 0) \\
\Phi_2^{(l=1)}(z,\bar{z}; 0)
\end{array}\right)=
\left(\begin{array}{c} \frac{z-z_0}{|z-z_0|} \\
0\end{array}\right)\cdot e^{{1\over 2}u(|z-z_0|;\,\, 0)} \qquad .
\]
\[
\bigtriangleup
\log[|z-z_0|^2+|q_0|^2]=\frac{4|q_0|^2}{(|z-z_0|^2+|q_0|^2)^2}
\qquad ,
\]
however, tends to zero for very large $|q_0|$ and the solution of
(\ref{eq:for2c}) becomes: $u(|z-z_0|; |q_0|>>1)\simeq 0$. Therefore,
\[
\left(\begin{array}{c} \Phi_1^{(l=1)}(z,\bar{z}; |q_0|>>1) \\
\Phi_2^{(l=1)}(z,\bar{z}; |q_0|>>1)
\end{array}\right)=\frac{1}{\sqrt{\left|z-z_0\right|^2+\left|q_0\right|^2}}\cdot
\left(\begin{array}{c} z-z_0 \\ q_0\end{array}\right)
\]
is precisely the field profile of the lump centered at $z_0$ with
radius $|q_0|$ and topological charge 1 in the planar ${\rm CP}^1$
model. The Higgs fields spreads over the vacuum manifold $S^3$ for
very large $|q_0|$ whereas the ANO profiles are found for very small
$|q_0|$. Self-dual semi-local topological solitons interpolate
between self-dual ANO vortices and ${\rm CP}^1$-lumps when $|q_0|$
varies between 0 and $\infty$.

\subsection*{Self-dual semi-local topological solitons with spherical symmetry}

Among the topological solitons, the simplest ones are those in which
all the roots of both $P_l(z)=(z-z_0)^l$ and $Q_{m}(z)=q_m(z-z_0)^m$
are at the same point. Let us focus on this case by considering the
spherically symmetric ansatz around this point chosen at the origin.
If $r=+\sqrt{x_1^2+x_2^2}$, for configurations of the form
\[
\Phi(x_1,x_2) = \left(\begin{array}{c}f(r) e^{il\theta}\\
|h(r)|e^{i(\lambda+ m\theta)}\end{array}\right) \quad , \quad
f(r)=\frac{r^{(l-m)}}{\sqrt{r^{2(l-m)}+|q_m|^2}}\cdot e^{{1\over
2}u(r;\,\, q_m)}
\]
\[
A_i(x_1,x_2)=-l \varepsilon_{ij} \frac{\alpha(r)}{r^2}x_j \qquad ,
\qquad h(r)=\frac{q_m}{\sqrt{r^{2(l-m)}+|q_m|^2}}\cdot e^{{1\over
2}u(r;\,\, q_m)} \qquad ,
\]
the first-order PDE's reduce to a system of nonlinear ordinary
differential equations:
\begin{eqnarray}
{1\over r} {d \alpha \over d r}&=& - \frac{1}{2 l} (f^2(r)+|h(r)|^2-1)\label{eq:sem1}\\
{d f\over d r}&=&  \frac{l}{r} f(r)[1-\alpha(r)] \label{eq:sem2} \\
{d |h|\over dr}&=&{l\over r}|h|(r)({m\over l}-\alpha(r))
\label{eq:sem3}
\end{eqnarray}
to be solved together with the boundary conditions
\begin{eqnarray}
&&{\displaystyle \lim_{r\rightarrow\infty}} f(r) = 1 \hspace{1cm} ,
\hspace{1cm} {\displaystyle \lim_{r\rightarrow\infty}}h(r) = 0
\hspace{1cm},\hspace{1cm} {\displaystyle
\lim_{r\rightarrow\infty}}\alpha(r) = 1 \nonumber \\ &&f(0) =0
 \hspace{1.5cm}
, \hspace{1.3cm}|h(0)|=|h_0|
\delta_{m,0}\hspace{0.6cm},\hspace{0.8cm}\alpha(0)=0
\label{eq:sembc}
\end{eqnarray}
required by energy finiteness plus regularity at the center of the
vortex. Let us recall that the boundary conditions at infinity also
require that $m {\rm be}<l$. The magnetic field and the energy
density of the spherically symmetric vortices in terms of the field
profiles $f(r)$, $\alpha(r)$ are:
\[
B(r)={l\over 2r}\frac{d\alpha}{dr}
\]
\[
 {\cal E}(r)={1\over 8}({1\over
l^2}+1)(1-f^2(r)-|h(r)|^2)^2+{l^2f^2(r)\over
r^2}(1-\alpha(r))^2+\frac{l^2|h(r)|^2}{r^2}({m\over l}-\alpha(r))^2
\qquad  .
\]

\subsection*{Semi-local strings with a quantum of magnetic flux}

We now go on to the most elementary solutions that carry a quantum
of magnetic flux, or, $l=1=m+1$ . We follow the procedure developed
in \cite{VeSh} to solve the non-linear ODE system
(\ref{eq:sem1})-(\ref{eq:sem2})-(\ref{eq:sem3}) with boundary
conditions (\ref{eq:sembc}). First, we consider small values of $r$
and in the first-order differential equations we test the power
series
\begin{eqnarray}
f(r)&\equiv& f_1\cdot r+f_2\cdot r^2+f_3\cdot r^3+f_4 r^4+ \cdots \label{eq:sdex1}\\
\alpha(r) & \equiv & \alpha_1\cdot r+\alpha_2\cdot r^2+\alpha_3\cdot
r^3+\alpha_4\cdot r^4+ \cdots \label{eq:sdex2}\\ h(r)&\equiv&
h_0+h_1\cdot r+h_2\cdot r^2+h_3\cdot r^3+ h_4\cdot r^4+\cdots \qquad
, \label{eq:sdex3}
\end{eqnarray}
where $f_j$ and $\alpha_j$, $j=1,2,3, \cdots $, are real, whereas
$h_j$, $j=0,1,2, \cdots $, are complex coefficients. The coupled
first-order ODE's are solved at this limit by
(\ref{eq:sdex1})-(\ref{eq:sdex2})-(\ref{eq:sdex3}) if
\begin{eqnarray*}
f(r) & \simeq & f_1\cdot r  +\frac{f_1}{8} (|h_0|^2-1)\cdot  r^3 +
\frac{f_1}{128} \left[ (|h_0|^2-1)(2|h_0|^2-1)+4 f_1^2 \right]\cdot
r^5
+\dots \\
\alpha(r) & \simeq &  \frac{1}{4}(1-|h_{0}|^2)\cdot r^2- \left[
\frac{1}{32} |h_0|^2 (|h_0|^2-1)+\frac{1}{8} f_1^2 \right]\cdot r^4 - \\
&-&  \left[ \frac{1}{768} |h_0|^2
(|h_0|^2-1)(3|h_0|^2-2)+\frac{1}{192} f_1^2 (5|h_0|^2-4)
\right]\cdot r^6
+\dots \\
h(r) & \simeq & h_0 + \frac{h_0}{8}(|h_0|^2-1)\cdot r^2+
\frac{h_0}{128}  \left[(|h_0|^2-1)(2|h_0|^2-1)+4 f_1^2\right]\cdot
r^4 +\dots \qquad .
\end{eqnarray*}
We stress that
\[
h_0=\frac{q_0}{|q_0|}\cdot e^{{1\over 2}u(0;\,\, q_0)}
\]
is determined by the behavior of the solution at the origin. After
that, only a free parameter, $f_1$, is left in the exact solution
near the origin. Second, a numerical scheme is implemented by
setting a boundary condition at a non-singular point of the ODE
system, which is obtained from the power series for a small value of
$r$ ($r=0.001$ in our case). This scheme prompts a shooting
procedure by varying $f_1$, where the correct asymptotic behavior of
the solutions is obtained by setting an optimal value for $f_1$ for
a given value of $h_0$. Finally, the first-order ODE system is
solved for large $r$ by means of a power series in ${1\over r}$:
\[
f(r) = \sum_{j=0} f^j\cdot r^{-j} \hspace{0.5cm} , \hspace{0.5cm}
\alpha(r) = \sum_{j=0} \alpha^j\cdot r^{-j} \hspace{0.5cm} ,
\hspace{0.5cm}h(r) = \sum_{j=1} h^j\cdot r^{-j} \quad \qquad ,
\]
with the result that
\begin{eqnarray*}
f(r) & \simeq & 1-\frac{|h^1|^2}{2}\cdot  r^{-2} + (-2|h^1|^2+
\frac{3}{8}|h^1|^4)\cdot
r^{-4}+|h^1|^2(-32+5|h^1|^2-\frac{5}{16}|h^1|^4)\cdot r^{-6}+\\&+&
|h^1|^2(-1152+158|h^1|^2-\frac{35}{4}|h^1|^4+\frac{35}{128}|h^1|^6)\cdot r^{-8}+\dots \\
\alpha(r) & \simeq &  1- |h^1|^2\cdot r^{-2}+|h^1|^2 (
-8+|h^1|^2)\cdot r^{-4} +|h^1|^2(-192+24|h^1|^2-|h^1|^4)\cdot r^{-6} \\
&& + |h^1|^2 (9216+1120|h^1|^2-48|h^1|^4+|h^1|^6)\cdot r^{-8}
+\dots \\
|h(r)| & \simeq & |h|^1\cdot r^{-1}+-\frac{|h^1|^3}{2}\cdot
r^{-3}+|h^1|^3(-2+{3\over 8}|h^1|^2)\cdot r^{-5}+\\&+&
|h^1|^3(-32+5|h^1|^2-\frac{5}{16}|h^1|^4)\cdot r^{-7} +\dots \qquad
.
\end{eqnarray*}
Again, only one free parameter, $h^1$, is left. The value of $h^1$
is fixed by demanding continuity of the solution at intermediate
distances ($r=15$ in our case) obtained by gluing the short-$r$ and
large-$r$ approximations. In particular, this has the important
implication that
\[
|h_0|=0 \qquad \qquad  \Rightarrow \qquad \qquad |h^1|=0 \qquad
\qquad ,
\]
linking the null value of $|h_0|$, which gives the embedded ANO
vortex, with the null value of the constant $|h^1|$ setting the
behavior of the solution for very large $r$. Another important
remark is that the large $r$ behavior of self-dual semi-local
defects differs from the large $r$ behavior of self-dual ANO
vortices that decay exponentially.

The following Figures show the results obtained with this procedure
for several values of $h_0$. Note that $h_0=\frac{q_0}{|q_0|}\cdot
e^{{1\over 2}u(0; q_0)}$, $u(0; 0)=-\infty$, $u(0;\infty)=0$,
$h_0=0$ for the ANO vortices, and $h_0=1$ for the ${\rm
CP}^1$-lumps. It is observed in the graphics that the field profiles
reach their vacuum values at distances of the order of $r=15$.
Consequently, almost identical numerical solutions would be
generated by sewing the numerical and the asymptotic solutions
together at $r$ greater than $15$.

\begin{figure}[htdp]
\centerline{\includegraphics[height=3.5cm]{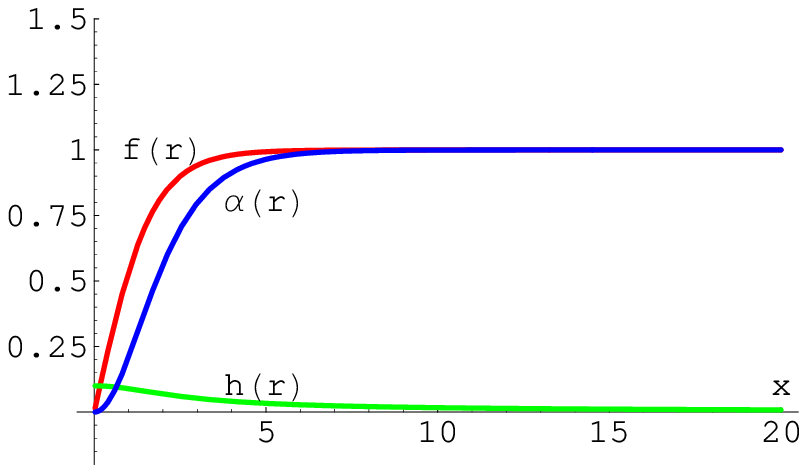}\hspace{0.3cm}
\includegraphics[height=3.5cm]{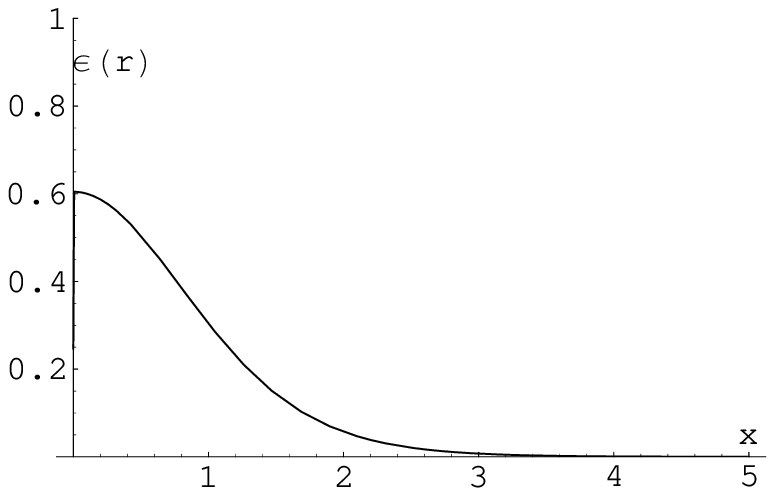}\hspace{0.3cm}
\includegraphics[height=3.5cm]{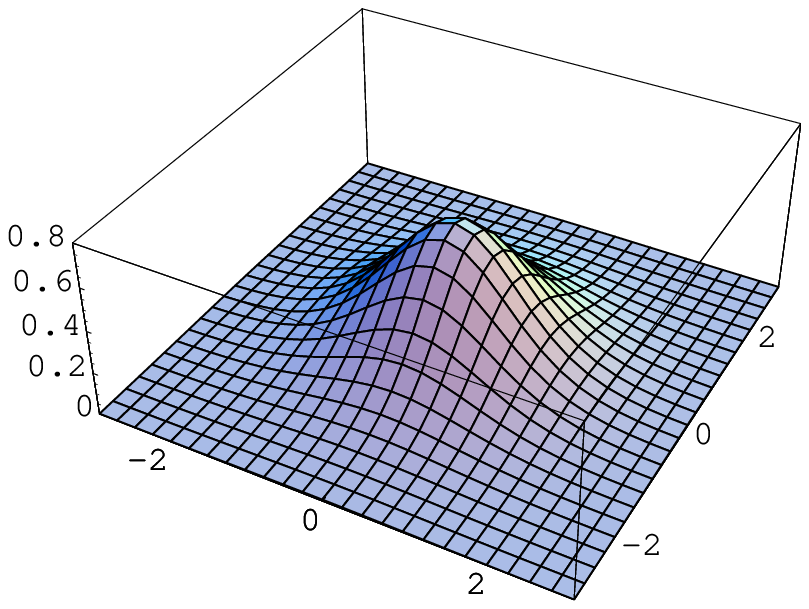}}
\centerline{\small \textit{a) Functions $f(r)$, $h(r)$ and
$\alpha(r)$ and b), c) Energy density for $h_0=0.1$}} \vspace{0.8cm}
\centerline{\includegraphics[height=3.5cm]{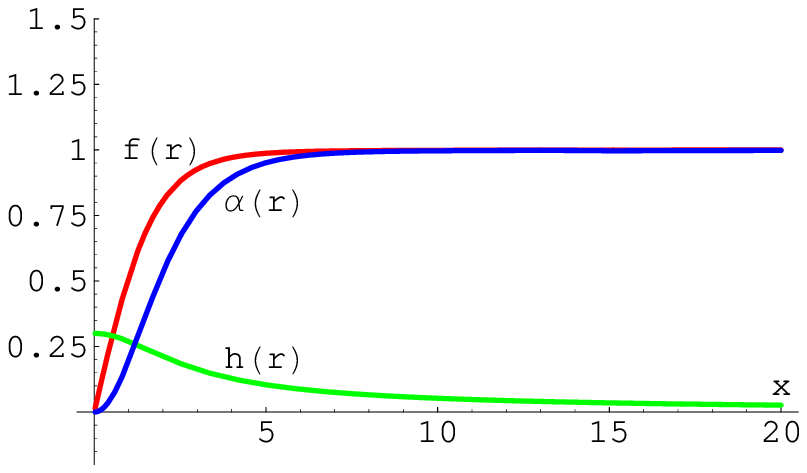}\hspace{0.3cm}
\includegraphics[height=3.5cm]{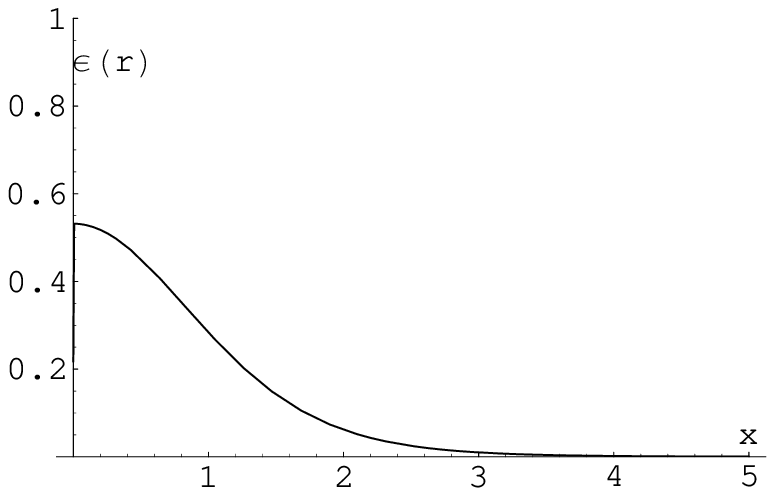}\hspace{0.3cm}
\includegraphics[height=3.5cm]{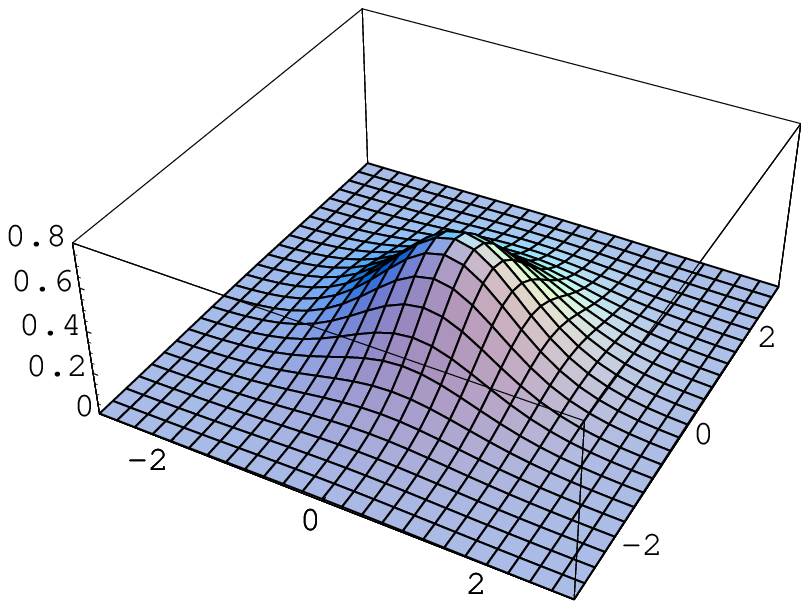}}
\centerline{\small \textit{a) Functions $f(r)$, $h(r)$ and
$\alpha(r)$ and b), c) Energy density for $h_0=0.3$}} \vspace{0.8cm}
\end{figure}

\begin{figure}[htdp]
\centerline{\includegraphics[height=3.5cm]{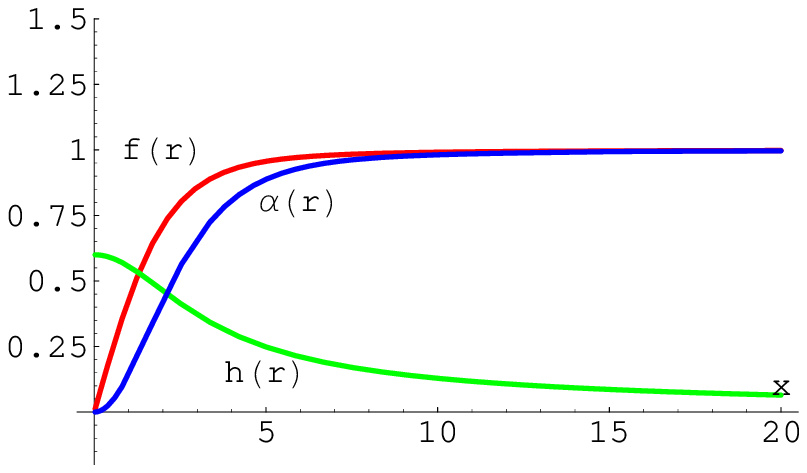}\hspace{0.3cm}
\includegraphics[height=3.5cm]{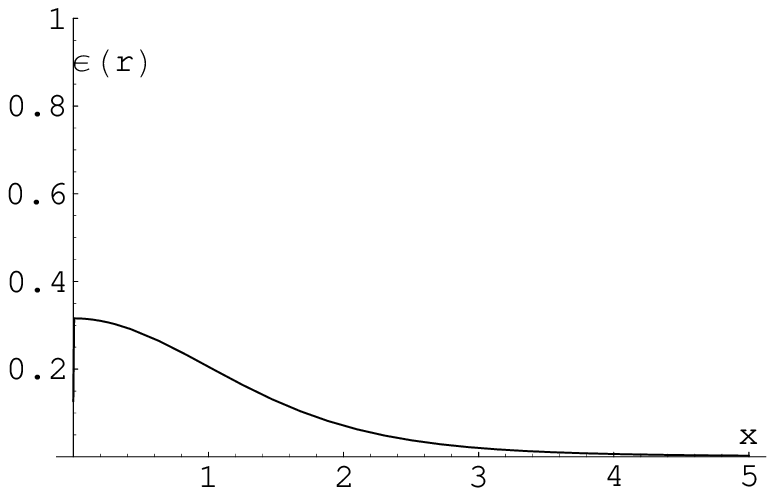}\hspace{0.3cm}
\includegraphics[height=3.5cm]{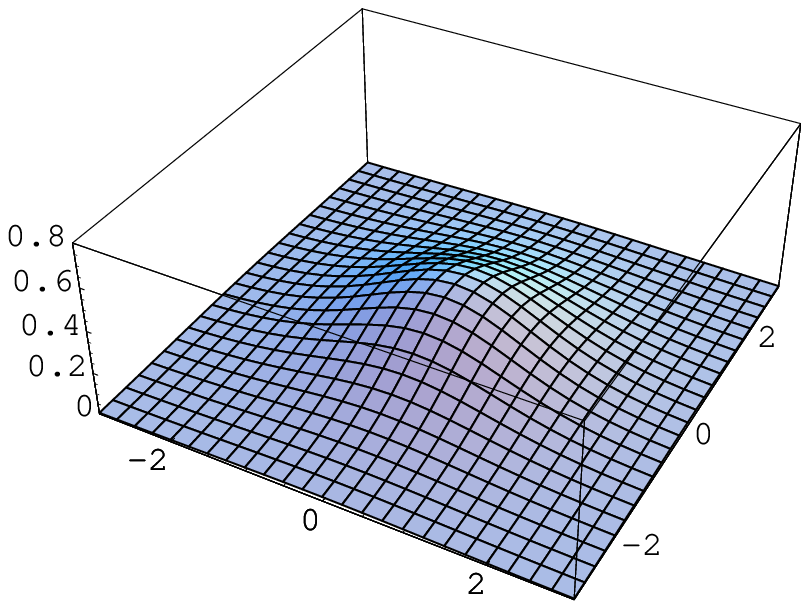}}
\centerline{\small \textit{a) Functions $f(r)$, $h(r)$ and
$\alpha(r)$ and b), c) Energy density for $h_0=0.6$}} \vspace{0.8cm}
\centerline{\includegraphics[height=3.5cm]{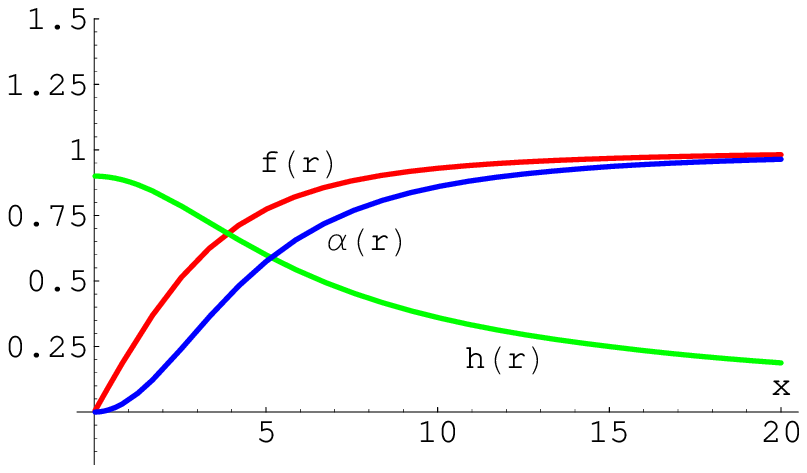}\hspace{0.3cm}
\includegraphics[height=3.5cm]{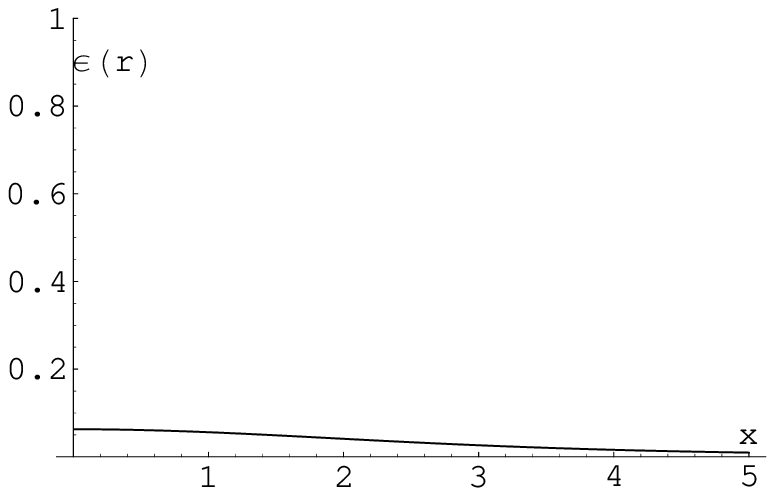}\hspace{0.3cm}
\includegraphics[height=3.5cm]{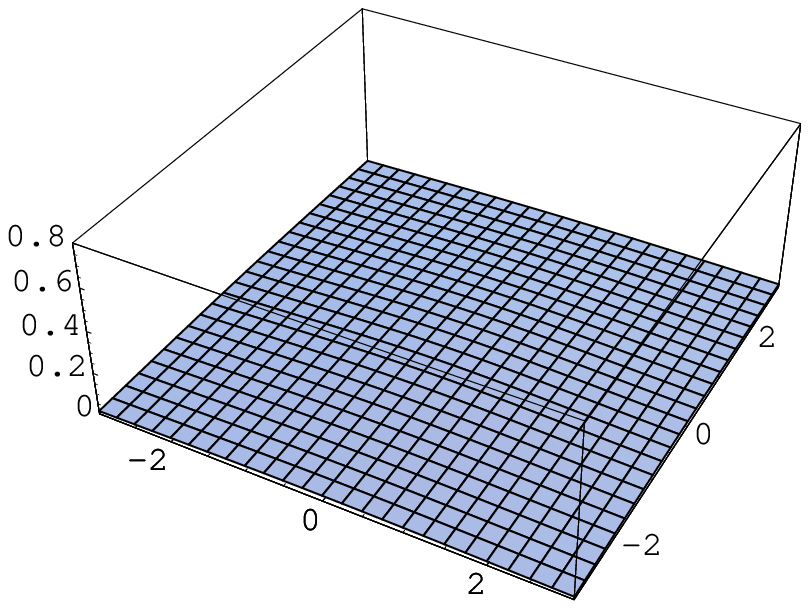}}
\centerline{\small \textit{a) Functions $f(r)$, $h(r)$ and
$\alpha(r)$ and b), c) Energy density for $h_0=0.9$}} \caption{Field
Profiles and Energy Densities for Semi-local Topological Defects}
\end{figure}

\section{One-loop correction to the masses of \\semi-local self-dual topological solitons}

\subsection*{Casimir energies}
The starting point for computing the quantum corrections to the
soliton mass is the evaluation of the zero-point oscillations of the
fields around both the soliton and the vacuum. The measurement of
the zero-point soliton energy with respect to the vacuum energy
shows a strong analogy with the Casimir effect, where the plates are
substituted by the soliton profile.

Let us first consider small fluctuations around the semi-local
soliton by writing:
\[
\Phi(\vec{x})=S(\vec{x})+\delta S(x_0,\vec{x}) \hspace{1.5cm} ,
\hspace{1.5cm} A_j(\vec{x})=V_j(\vec{x})+\delta a_j(x_0,\vec{x})
\qquad .
\]
By $S(\vec{x})$ and $V_k(\vec{x})$ we respectively denote the scalar
and vector fields of the topological soliton solution, whereas
$\delta S(x_0,\vec{x})$ and $\delta a_j(x_0,\vec{x})$ calibrate the
small deviations of the bosonic fields with respect to the classical
solution. Of course, not all fluctuations are physically relevant,
and we should avoid pure gauge deformations. In order to do so, we
impose the Weyl/background gauge condition:
\[
A_0(x_0,\vec{x})=0 \qquad \qquad , \qquad \qquad \partial_j\delta
a_j(x_0,\vec{x})+ \frac{i}{2}(S^\dagger(\vec{x}) \delta
S(x_0,\vec{x})-\delta S^\dagger(x_0,\vec{x}) S(\vec{x}))=0 \qquad .
\]
Under these constraints, the ground state energy in the topological
sectors up to ${\cal O}(\delta^3)$ order (one-loop) is:
\begin{equation}
H^{(2)}={v^2\over 2}\int \,
d^2x\left\{\frac{\partial\delta\xi^T}{\partial
x_0}\frac{\partial\delta\xi}{\partial x_0}+\delta \xi^T K \delta\xi+
\delta\chi^* K^G \delta\chi\right\} \label{eq:h2} \qquad ,
\end{equation}
where the bosonic fluctuations are arranged in the column vector
\[
\delta\xi(x_0,\vec{x})=\left(\begin{array}{c} \delta a_1 (x_0,\vec{x}) \\
\delta a_2 (x_0,\vec{x}) \\ \delta S_1^1(x_0,\vec{x}) \\ \delta
S_1^2(x_0,\vec{x}) \\\delta S_2^1(x_0,\vec{x}) \\ \delta
S_2^2(x_0,\vec{x})  \end{array}\right) \qquad \qquad ,
\]
and $\delta \chi(\vec{x})$ corresponds to ghost fluctuations. $K$
and $K^G$ are second-order differential operators. $K$ is
matrix-valued and fairly complicated
\begin{displaymath}
K=\left(\begin{array}{cccccc}
A&0&-2\nabla_1 S_1^2 &2\nabla_1 S_1^1 & -2\nabla_1 S_2^2 &2\nabla_1 S_2^1  \\
0&A&-2\nabla_2 S_1^2 & 2\nabla_2 S_1^1 & -2\nabla_2 S_2^2 & 2\nabla_2 S_2^1 \\
-2\nabla_1 S_1^2 &-2\nabla_2 S_1^2 &B&-2 V_k\partial_k&S_1^1S_2^1+S_1^2S_2^2 &S_1^1S_2^2-S_1^2S_2^1 \\
2\nabla_1 S_1^1 & 2\nabla_2 S_1^1&2 V_k \partial_k&B&-S_1^1S_2^2+S_1^2S_2^1 & S_1^1S_2^1+S_1^2S_2^2 \\
-2\nabla_1 S_2^2 &-2\nabla_2 S_2^2  &S_1^1S_2^1+S_1^2S_2^2  & -S_1^1S_2^2+S_1^2S_2^1 &C&-2 V_k\partial_k\\
2\nabla_1 S_2^1 & 2\nabla_2 S_2^1 &S_1^1S_2^2-S_1^2S_2^1  &
S_1^1S_2^1+S_1^2S_2^2 &2 V_k\partial_k&C
\end{array}\right) \qquad , \label{eq:bsfats1}\nonumber
\end{displaymath}
and $K^G$ is the scalar differential operator:
\begin{displaymath}
K^G=-\partial_k\partial_k+|S_1|^2+|S_2|^2 \quad .\nonumber
\end{displaymath}
In all these formulas, the notation is
\[
S(\vec{x})=\left(\begin{array}{c}S_1(\vec{x})\\
S_2(\vec{x})\end{array}\right)=\left(\begin{array}{c}S_1^1(\vec{x})+i S_1^2(\vec{x}\\
S_2^1(\vec{x})+i S_2^2(\vec{x})\end{array}\right) \qquad \qquad ,
\]
and analogously for $\delta S$. The covariant derivatives are
$\nabla_j S_M^a=\partial_j S_M^a+\varepsilon^{ab} V_jS_M^b$, and the
differential operators in the diagonal of $K$ take the form
\begin{eqnarray*}
A&=&-\partial_k \partial_k+|S_1|^2+|S_2|^2 \qquad , \qquad
B=-\partial_k \partial_k+\frac{1}{2}(3|S_1|^2+|S_2|^2+2V_k V_k-1) \quad ,\\
C&=&-\partial_k\partial_k+\frac{1}{2}(|S_1|^2+3|S_2|^2+2V_k
V_k-1)\,\, .
\end{eqnarray*}

The energy due to the zero point fluctuations around the vacuum must
be subtracted. As is fairly obvious, the Hamiltonian for the vacuum
fluctuations has exactly the same form as (\ref{eq:h2}) for the
vacuum values: $S_1=1, S_2=0, V_1=V_2=0$. We shall denote by $K_0$
and $K^G_0$ the operators $K$ and $K^G$ in the vacuum background.

When all the fluctuation modes are unoccupied, we formally obtain
the contribution to the semi-local topological soliton ground state
energy as a difference between two \lq\lq super-traces" of
differential operators acting on column vectors of $L^2({\mathbb
R}^2)$ functions. The first super-trace comes from (the star
stresses the fact that zero modes do not contribute to the trace)
oscillations of the fields around the soliton,
\[
\bigtriangleup E_{\rm TS}= \frac{\hbar m}{2}\cdot\left({\rm
Tr}^*K^{{1\over 2}} -{\rm Tr}^*\left(K^G\right)^{{1\over
2}}\right)=\frac{\hbar m}{2}\cdot{\rm STr}^*K^{{1\over 2}}\qquad ,
\]
and the second one accounts for vacuum fluctuations of the fields:
\[
\bigtriangleup E_0=\frac{\hbar m}{2}\cdot\left({\rm Tr}K_0^{{1\over
2}}-{\rm Tr}\left(K_0^G\right)^{{1\over 2}}\right)={\rm
STr}K_0^{{1\over 2}} \qquad .
\]
Therefore,
\begin{equation}
\bigtriangleup M_{\rm TS}^C=\bigtriangleup E_{\rm TS}-\bigtriangleup
E_0 ={\hbar m\over 2}\left({\rm STr^*}\, K^{{1\over 2}}- {\rm STr}\,
K_0^{{1\over 2}}\right) \label{eq:sltsc}
\end{equation}
formally measures the semi-local topological soliton Casimir energy.

Before proceeding, a further explanation of how we choose small
fluctuations belonging to $L^2({\mathbb R}^2)$ deserves a pause.
Following the conventional QFT approach, we put the system in a very
large but finite two-dimensional box and impose periodic boundary
conditions on $\delta\xi(x_0,\vec{x})$ and $\delta\chi(\vec{x})$. In
Appendices II, III, and IV of Reference \cite{AGGMMT} it is shown
that taking the infinite area limit at the end leaves no remnants
for kinks and self-dual vortices. Thus, this procedure uses
invisible boundary conditions in such a way that, at intermediate
stages, one works on circles or genus one Riemann Surfaces.
Nevertheless, the rapid (exponential) decay of the Higgs field to
its vacuum value in these cases ensures that, starting from large
lengths or areas, there will be a very small dependence of the
results on the size. Except for the embedded ANO vortices, all the
Higgs fields of the semi-local self-dual topological solitons decay
to their vacuum values as ${1\over r^k}$ for some positive $k$.
Therefore, we expect a more significant dependence on the size of
the box for these less concentrated solitons.

We might try topological boundary conditions like those used in
Reference \cite{Reb1} for the bosonic and supersymmetric kink. In
this planar gauge theoretical setting, the analogous form of the
anti-periodic boundary conditions would be:
\[
\delta\Phi(x_1,x_2)=e^{i\int_{P_0}^{P_{mL}}\,
[V_k(\vec{x}^\prime)dx_k^\prime\, ]}\cdot \delta\Phi(x_1+mL,x_2+mL)
\qquad .
\]
The line integral is along a path starting at the point
$P_0=(x_1,x_2)$, ending at $P_{mL}=(x_1+mL,x_2+mL)$, and passing
through regions far away from the vortex core. These twisted
boundary conditions are also invisible in the sense that no boundary
is introduced. Rather, the fact that we are dealing with a
non-trivial line bundle of first Chern class equal to one over a
genus one Riemann surface is taken into account. In a purely bosonic
framework, however, the twisted boundary conditions, like periodic
boundary conditions, will not leave any mark at the infinite area
limit.

\subsection*{Counter-term energies}

The Casimir energy of the previous Section is of order $\hbar$, and
this is also the order of the counter-terms found in Section 2.
Thus, the one-loop semi-local topological soliton mass also receives
contributions from the counter-terms for scalar and vector fields.
At the self-dual limit $\kappa^2=1$, these contributions are
\[
\Delta M_{c.t.}^S=\frac{\hbar\, m}{2}[4I(1)+I(0)]\,\int \, d^2x
\,(1-|S_1|^2-|S_2|^2) \,\,\, , \,\,\, \Delta M_{c.t.}^A=-\hbar \, m
\, [I(1)+I(0)]\int \, d^2x \, V_kV_k \quad .
\]
We now reshuffle the sum of these two quantities into two pieces,
respectively proportional to $I(1)$ and $I(0)$:
\begin{equation}
\bigtriangleup M_{\rm c.t.}^{\rm I(1)}= \frac{\hbar\, m}{2}\,I(1)
\,\Sigma^{(1)}(S,V_k) \qquad , \qquad  \bigtriangleup M_{\rm
c.t.}^{\rm I(0)}= \frac{\hbar\, m}{2}\,I(0)
\,\Sigma^{(0)}(S,V_k)\label{eq:renes} \qquad ,
\end{equation}
where
\[
\Sigma^{(1)}(S,V_k)=4\int \, d^2x \, (1-|S_1|^2-|S_2|^2-{1\over
2}V_kV_k )\,\, , \,\,  \Sigma^{(0)}(S,V_k)= \int \, d^2x
\,\left(1-|S_1|^2-|S_2|^2-2V_kV_k\right) \,.
\]
The total contribution to the one-loop mass shifts from the mass
renormalization counter-terms is:
\[
\bigtriangleup M_{\rm c.t.}=\bigtriangleup M_{\rm c.t.}^{\rm
I(1)}+\bigtriangleup M_{\rm c.t.}^{\rm I(0)}  \qquad .
\]

\subsection*{Deformation of the first-order equations}

It follows from the classical degeneracy of vortices that there are
some static deformations of a topological solution that do not cost
energy at the tree level: those that give rise to a new self-dual
soliton. Therefore, the dimension of the kernel of the operator $K$
ruling the bosonic small fluctuations around the topological soliton
should be equal to the dimension of the moduli space of self-dual
semi-local soliton solutions. To check this, we proceed as in the
previous Section by expanding the bosonic fields around the
solution, but this time we consider only static small fluctuations:
\[
\Phi(\vec{x})=S(\vec{x})+\delta S(\vec{x}) \hspace{1.5cm} ,
\hspace{1.5cm} A_j(\vec{x})=V_j(\vec{x})+\delta a_j(\vec{x})\qquad .
\]
The modified fields are still solutions of the first-order equations
if (\ref{eq:slsdm1}) and (\ref{eq:slsdm2}) are satisfied:
\begin{equation}
F_{12}={1\over 2}(1-\Phi^\dagger \Phi)\Leftrightarrow
-\partial_2\delta a_1+\partial_1\delta a_2+\frac{1}{2}(S^\dagger
\delta S+\delta S^\dagger S)=0 \label{eq:slsdm1} \qquad ,
\end{equation}
and
\begin{equation}
D_1\Phi+iD_2\Phi=0\Leftrightarrow (\nabla_1+i\nabla_2)\delta
S-i(\delta a_1+i\delta a_2) S=0 \label{eq:slsdm2} \qquad ,
\end{equation}
where $\nabla_k=\partial_k-i V_k$. Also as in the previous Section,
in order to avoid pure gauge fluctuations we set the static
background gauge:
\[
\partial_j\delta a_j(\vec{x})+\frac{i}{2}(S^\dagger \delta S-\delta S^\dagger S)=0 \quad .
\]

Thus, the tangent space to the moduli space of self-dual vortices is
the kernel of the first-order deformation operator ${\cal D}$:
\[
{\cal D}= \left(\begin{array}{cccccc}
-\partial_2 &\partial_1 & S_1^1 & S_1^2&S_2^1&S_2^2 \\
 -\partial_1 & -\partial_2 & -S_1^2 & S_1^1&-S_2^2&S_2^1\\
S_1^1 & -S_1^2 & -\partial_2+V_1 & -\partial_1-V_2 &0&0\\
S_1^2 &S_1^1&\partial_1+V_2 & -\partial_2+V_1&0&0\\
S_2^1 & -S_2^2&0&0 & -\partial_2+V_1 & -\partial_1-V_2\\
S_2^2 &S_2^1&0&0&\partial_1+V_2 & -\partial_2+V_1\\
 \end{array}\right) \qquad .
\]
One easily checks that $K={\cal D}^\dagger{\cal D}$, and it is also
possible to prove that the Kernel of ${\cal D}^\dagger$ is empty
\cite{GG99}. Thus, $K$ and ${\cal D}$ have the same kernel and the
dimension of that kernel is the dimension of the classical moduli
space: $4l$.

\subsection*{Zeta-function regularization}

The zero-point energies for the topological soliton and the vacuum
are formally super-traces: the differences of traces of differential
operators. Such traces are divergent quantities. The standard
procedure for dealing with this delicate point is understanding
these traces as the generalized zeta functions of the corresponding
differential operators; see \cite{Vass}-\cite{Kirsten}-\cite{Gil}.
Given a differential operator $A$ acting on an $L^2$ space of
functions, the corresponding generalized zeta function is:
\[
\zeta_A(s)=\sum_{{\rm Spec}\,A}\,\lambda_n^{-s} \qquad \qquad ,
\qquad \qquad s\in{\mathbb C} \qquad ,
\]
where $\lambda_n$ are the eigenvalues of $A$.  We shall regularize
the zero-point energies in the form:
\[
\Delta E_{TS}(s)=\frac{\hbar\mu}{2}\left({\mu^2\over
m^2}\right)^s\left\{\zeta_{K}(s)-\zeta_{K^G}(s)\right\}\hspace{0.6cm},
\hspace{0.6cm} \Delta E_0(s)=\frac{\hbar\mu}{2}\left({\mu^2\over
m^2}\right)^s\left\{ \zeta_{K_0}(s)-\zeta_{K^G_0}(s)\right\} \quad ,
\]
i.e., by assigning to the meromorphic generalized zeta functions the
value obtained by analytic continuation at the point $s$ of the $s$
-complex plane. The physical limit is:
\[
\Delta M_{TS}^C=\lim_{s\rightarrow -\frac{1}{2}}\Delta
M_{TS}^C(s)=\lim_{s\rightarrow -\frac{1}{2}}\left(\Delta
E_{TS}(s)-\Delta E_0(s)\right).
\]
Because $K_0$ and $K^G_0$ are free Schr\"{o}dinger operators, their
zeta functions are well known \cite{Aai1, AGMT}:
\[
\zeta_{K_0}(s)=\frac{m^2L^2}{\pi}\cdot
\frac{\Gamma[s-1]}{\Gamma(s)}+ \frac{m^2L^2}{2\pi}\cdot
\frac{1}{(s-1)\Gamma(s)}\hspace{1cm} , \hspace{1cm}
\zeta_{K^G_0}(s)=\frac{m^2L^2}{4\pi}\cdot
\frac{\Gamma[s-1]}{\Gamma(s)}\quad ,
\]
where $\Gamma(s)$ and $\gamma[s,a]$ are the complete and incomplete
Euler Gamma functions, respectively.

The contribution from the mass counter-terms also involves divergent
quantities proportional to the integrals $I(1)$ and $I(0)$. To
regularize these integrals, we apply the residue theorem to
integrate $I(c)$ in the complex $k_0$-plane. On a square of area
$m^2L^2$ each integral becomes a infinite sum over discrete momenta:
\[
I(c)={1 \over 2}\int {d^2k\over (2 \pi)^2} {1\over \sqrt{\vec k
\cdot \vec k +c}}={1\over 2}{1\over m^2L^2}\sum_{\vec{k}\in{\mathbb
Z}^2}\frac{1}{\sqrt{\vec{k}\cdot\vec{k}+c}} \qquad .
\]
Accordingly, these integrals are the generalized zeta functions of
the Euclidean Klein-Gordon and Laplacian operators evaluated at
$s={1\over 2}$. In this way, we also define the mass renormalization
correction as a meromorphic function in the complex $s$-plane:
\[
\Delta M_{TS}^R(s) = {\hbar\over 2\mu L^2}\left({\mu^2\over
m^2}\right)^s \left(\zeta_{-\bigtriangleup+1} (s) \Sigma^{(1)}
(S,V_k)+\zeta_{-\bigtriangleup}(s) \Sigma^{(0)} (S,V_k)\right)
\qquad ,
\]
and take the physical limit:
\[
\Delta M_{TS}^R=\lim_{s\rightarrow{1\over 2}} \Delta M_{TS}^R(s)
\qquad \qquad .
\]
While $-\bigtriangleup+1$ is exactly $K^G_0$, and its zeta function
has been written above, calculation of $\zeta_\bigtriangleup(s)$ is
a bit tricky. From the partition function for the Laplacian, via the
Mellin transform, we have:
\[
\zeta_{-\bigtriangleup}(s)={m^2L^2\over
4\pi}\lim_{\varepsilon\rightarrow
0}{1\over\Gamma(s)}\left(\int_0^1\, d\beta \, \beta^{s-2}\,
e^{-\varepsilon\beta}+\int_1^\infty\, d\beta \, \beta^{s-2}\,
e^{-\varepsilon\beta}\right) \qquad ,
\]
or,
\[
\zeta_{-\bigtriangleup}(s)={m^2L^2\over
4\pi}\lim_{\varepsilon\rightarrow
0}{1\over\varepsilon^{s-1}\Gamma(s)}\left(\gamma[s-1,\varepsilon]+\Gamma[s-1,\varepsilon]\right)\qquad
.
\]
Because $\gamma[s-1,\varepsilon]{\footnotesize{\begin{array}{c}\\ \cong \\
\varepsilon\rightarrow
0\end{array}}}{\Large{\frac{\varepsilon^{s-1}}{s-1}-{\varepsilon^s\over
s}}}$ and $\Gamma[s-1,\varepsilon]{\footnotesize{\begin{array}{c}\\ \cong \\
\varepsilon\rightarrow 0\end{array}}}{\large{\Gamma[s-1]}}$ , we
obtain:
\[
\zeta_{-\bigtriangleup}(s)={m^2L^2\over 4\pi}\cdot{1\over
(s-1)\Gamma(s)} \qquad {\rm if}\quad {\rm Re}\,\, {\rm s}<1 \qquad
\qquad .
\]
We finally find:
\[
I(1)={1\over 2m^2L^2}\zeta_{-\bigtriangleup+1}({1\over 2})={1 \over
8 \pi} {\Gamma (-{1\over 2}) \over \Gamma({1\over 2})}=-{1\over
4\pi} \qquad , \qquad I(0)={1\over
2m^2L^2}\zeta_{-\bigtriangleup}({1\over
2})=-\frac{1}{4\pi\sqrt{\pi}} \qquad .
\]
Contrary to the kink cases, which are one-dimensional problems, a
finite answer is obtained in the regularized integrals via the
associated zeta functions. The reason is that in this
two-dimensional problem the physical limit $s={1\over 2}$ is not a
pole of the zeta functions and only finite renormalizations will be
necessary.

\section{The semi-local topological soliton heat kernel \\ and generalized zeta function}

Control of $\zeta_{K}(s)$ and $\zeta_{K^G}(s)$ is much more
difficult. A convenient way for dealing with the zeta functions of
differential operators acting on infinite dimensional spaces is by
means of heat kernel techniques. In this Section, we shall develop
this method, applied to our soliton operators.

\subsection*{The heat kernel of a differential operator. Seeley densities}
The heat equation kernel of a $N\times N$ matrix differential
operator of the general form
\[
{\cal K}={\cal K}_0 +Q_k(\vec{x})\partial_k+V(\vec{x})
\]
is the solution of the ${\cal K}$-heat equation
\[
\left(\frac{\partial}{\partial\beta}{\mathbb I}+{\cal K}
\right)K_{{\cal K}}(\vec{x},\vec{y};\beta )=0
\]
with initial condition: $K_{{\cal K}}(\vec{x},\vec{y};0)={\mathbb
I}\cdot \delta^{(2)}(\vec{x}-\vec{y})$. We are particularly
interested in the diagonal $\vec{x}=\vec{y}$ heat-kernel, because
the Mellin transform of the partition function
\[
{\rm Tr}\,e^{-\beta {\cal K}}={\rm tr}\int_{{\mathbb R}^2} \,
d^2\vec{x} \, K_{{\cal K}}(\vec{x},\vec{x};\beta)
\]
gives the generalized zeta function.

To find the kernel, one writes \cite{Ston}
\[
K_{{\cal K}}(\vec{x},\vec{y};\beta)=C_{\cal
K}(\vec{x},\vec{y};\beta)K_{{\cal K}_0}(\vec{x},\vec{y};\beta)
\qquad ,
\]
where ${\cal K}_0$ is the operator ${\cal K}$ very distant from the
origin, where $Q_k(\vec{x})$ and $V(\vec{x})$ take their asymptotic
constant vacuum values. $C_{\cal K}(\vec{x},\vec{y};\beta)$
satisfies the $N\times N$-matrix transfer equation
\[
\left\{ {\partial\over\partial\beta}{\mathbb
I}+{x_k-y_k\over\beta}(\partial_k{\mathbb I}-{1\over
2}Q_k)-\bigtriangleup{\mathbb I}+Q_k\partial_k+V
\rule[-0.1in]{0.in}{0.2in} \right\}C_{\cal
K}(\vec{x},\vec{y};\beta)=0
\]
and is the unit matrix $C_{\cal K}(\vec{x},\vec{y};0)={\mathbb I}$
at infinite temperature.

Solving the transfer equation as an inverse-temperature power series
expansion,
\[
C_{\cal K}(\vec{x},\vec{y};\beta)=\sum_{n=0}^\infty
c_n(\vec{x},\vec{y};{\cal K})\beta^n,
\]
the PDE equation becomes tantamount to the recurrence relation
between the densities $c_n(\vec{x},\vec{y};{\cal K})$:
\[
[n{\mathbb I}+(x_k-y_k)(\partial_k{\mathbb I}-{1\over
2}Q_k)]c_n(\vec{x},\vec{y};{\cal K}) =[\bigtriangleup{\mathbb I}
-Q_k\partial_k-V]c_{n-1}(\vec{x},\vec{y};{\cal K})\hspace{1cm} ,
\hspace{1cm} n\geq 1
\]
to be started from:  $c_0(\vec{x},\vec{y};{\cal K})={\mathbb I}$.
While it is easy to find the first diagonal density,
$c_1(\vec{x},\vec{x};{\cal K})=-V(\vec{x})$, the determination of
higher-order densities becomes more and more involved. To make the
problem more tractable, we introduce the following notation:
\[
{}^{(\alpha_1,\alpha_2)}C_n^{AB}(\vec{x})=\lim_{\vec{y}\rightarrow
\vec{x}} \frac{\partial^{\alpha_1+\alpha_2}
[c_n]_{AB}(\vec{x},\vec{y};{\cal K})}{\partial
x_1^{\alpha_1}\partial x_2^{\alpha_2}} \hspace{1cm} , \hspace{1cm}
[{c}_n]_{AB}(\vec{x},\vec{x};{\cal
K})={}^{(0,0)}C_n^{AB}(\vec{x})\qquad .
\]
Thus, at the $\vec{y}\rightarrow\vec{x}$ limit the recurrence
relations between densities and partial derivatives of densities can
be written in compact form:

\begin{eqnarray*} &&(k+\alpha_1+\alpha_2+1)
{}^{(\alpha_1,\alpha_2)}C_{k+1}^{AB}(\vec{x})=
{}^{(\alpha_1+2,\alpha_2)}C_{k}^{AB}(\vec{x})+
{}^{(\alpha_1,\alpha_2+2)}C_{k}^{AB}(\vec{x})-  \\
&&-\sum_{d=1}^N \sum_{r=0}^{\alpha_1}\sum_{t=0}^{\alpha_2} {\alpha_1
\choose r} {\alpha_2 \choose t} \left[ \frac{\partial^{r+t}
Q^{AD}_1}{\partial x_1^r\partial x_2^t}
{}^{(\alpha_1-r+1,\alpha_2-t)}C_{k}^{DB}(\vec{x})\right.+\\&&+\left.
\frac{\partial^{r+t} Q^{AD}_2}{\partial x_1^r\partial x_2^t}
{}^{(\alpha_1-r,\alpha_2-t+1)}C_{k}^{DB}(\vec{x}) \right]+  \\&&
+\frac{1}{2}\sum_{d=1}^N
\sum_{r=0}^{\alpha_1-1}\sum_{t=0}^{\alpha_2} \alpha_1{\alpha_1-1
\choose r} {\alpha_2 \choose t}  \frac{\partial^{r+t}
Q^{AD}_1}{\partial x_1^r\partial x_2^t}
{}^{(\alpha_1-1-r,\alpha_2-t)}C_{k+1}^{DB}(\vec{x})+ \\
&&+\frac{1}{2}\sum_{d=1}^N
\sum_{r=0}^{\alpha_2-1}\sum_{t=0}^{\alpha_1} \alpha_2{\alpha_2-1
\choose r} {\alpha_1 \choose t}  \frac{\partial^{r+t}
Q^{AD}_2}{\partial x_1^t\partial x_2^r}
{}^{(\alpha_1-t,\alpha_2-1-r)}C_{k+1}^{DB}(\vec{x})-
\\&&-\sum_{d=1}^N
\sum_{r=0}^{\alpha_2}\sum_{t=0}^{\alpha_1}{\alpha_1 \choose
t}{\alpha_2\choose r}
 \frac{\partial^{r+t}
V^{AD}}{\partial x_1^t\partial x_2^r}
{}^{(\alpha_1-t,\alpha_2-r)}C_k^{DB}(\vec{x})
\end{eqnarray*}
to be solved starting from
\[
c_0(\vec{x},\vec{x};{\cal K})={\mathbb I} \Rightarrow\left\{
\begin{array}{c} {}^{(\alpha,\beta)}C_0^{AB}(\vec{x})=0 , \, {\rm
if}\,
\alpha \neq 0, {{\rm and}/{\rm or}}\, \beta \neq 0 \\
{}^{(0,0)}C_0^{AA}(\vec{x})=1 , \, A=1,2, \cdots ,N
\end{array}\right. \qquad .
\]
\subsection*{The Mellin transform of the asymptotic expansion}

We must now deal with the cases $N=6$ and $N=1$, respectively for
the operators $K$ and $K^G$. A good approximation to the generalized
zeta functions of both operators is obtained from the Mellin
transform \cite{Gil}
\[
\zeta_{K}(s)={1\over\Gamma(s)}\int_0^\infty \, d\beta \, \beta^{s-1}
\, {\rm Tr} \, e^{-\beta K} \qquad , \qquad
\zeta_{K^G}(s)={1\over\Gamma(s)}\int_0^\infty \, d\beta \,
\beta^{s-1} \, {\rm Tr} \, e^{-\beta K^G}
\]
applied to the high-temperature expansion of the partition functions
\begin{eqnarray*}
{\rm Tr}e^{-\beta K}&=&{1\over 4\pi\beta}
\sum_{n=0}^\infty\beta^n\cdot\left(e^{-\beta}\sum_{A=1}^4
[c_n^I(K)]_{AA}+ \sum_{A=5}^6 [c^O_n(K)]_{AA}\right)\\
{\rm Tr}e^{-\beta K^{\rm G}}&=&{e^{-\beta}\over 4\pi\beta}
\sum_{n=0}^\infty\beta^n c_n(K^{\rm G}) \qquad ,
\end{eqnarray*}
where
\begin{eqnarray*}
[c_n^I(K)]_{AA}&=&\int \, d^2x \,e^{-\beta}[c_n]_{AA}(\vec{x},\vec{x};K)\quad,\quad A=1,2,3,4\\
{[c_n^O(K)]}_{AA}&=&\int \, d^2x
\,[c_n]_{AA}(\vec{x},\vec{x};K)\quad,\quad A=5,6\quad , \quad
{[c_n(K^G)]}=\int \, d^2x \,[c_n](\vec{x},\vec{x};K^G) \,\, .
\end{eqnarray*}
The factor $e^{-\beta}$, which appears in front of $[c_n(K)]_{AA}$
for $A=1,2,3,4$, obeys the fact that the corresponding modes in
$\xi$ have one unit of mass, while the modes for $A=5,6$ are
massless.  The generalized zeta functions are thus divided as sums
of meromorphic -high-temperature regime- and entire -to be
neglected, low temperature regime- functions of $s$:

\begin{eqnarray*}
\zeta_{K}(s)&=&{1\over 4 \pi\Gamma(s)}\sum_{n=0}^\infty
\left\{\sum_{A=1}^4\int_0^1 \, d\beta \,
\beta^{s+n-2}[c_n^I(K)]_{AA}e^{-\beta}+\sum_{A=5}^6\int_0^1 \,
d\beta \,
\beta^{s+n-2}[c_n^O(K)]_{AA}\right\}\\&+&{1\over\Gamma(s)}\int_1^\infty
\, d\beta \, {\rm Tr}^* e^{-\beta K}\\&=&\sum_{n=0}^\infty
\left\{\sum_{A=1}^4[c_n^I(K)]_{AA}\frac{\gamma[s+n-1,1]}{4\pi\Gamma(s)}+
\sum_{A=5}^6[c_n^O(K)]_{AA}\frac{1}{4\pi
\Gamma(s)(s+n-1)}\right\}+{1\over\Gamma(s)}B_{K}(s)
\end{eqnarray*}
\begin{eqnarray*}
\zeta_{K^G}(s)&=&{1\over\Gamma(s)}\sum_{n=0}^\infty \int_0^1 \,
d\beta \,
\beta^{s+n-2}c_n(K^G)e^{-\beta}+{1\over\Gamma(s)}\int_1^\infty \,
d\beta \, {\rm Tr}^* e^{-\beta K^{G}}\\&=&\sum_{n=0}^\infty
c_n(K^G)\frac{\gamma[s+n-1,1]}{4\pi\Gamma(s)}+{1\over\Gamma(s)}B_{K^G}(s)
\qquad ,
\end{eqnarray*}
where $\gamma[s+n-1,1]$ are incomplete Euler Gamma functions. We
shall neglect the entire parts $B(K)$ and $B(K^G)$ and keep a finite
number of terms, $N_0$, in future use of these generalized zeta
functions for the regularization of ultraviolet divergences.

\subsection*{The high-temperature one-loop semi-local
vortex mass shift formula}

The contribution of the $c_1$ coefficients to the semi-local
topological soliton Casimir energy is
\[
\Delta M_{TS}^{(1)C} (s)  = {\hbar \over 2} \mu \left( {\mu^2 \over
m^2}\right)^s \left\{\left[\sum_{A=1}^4[c_1^I(K)]_{AA}-c_1(K^{\rm
G})\right] \cdot {\gamma[s,1] \over 4 \pi \Gamma(s)}+
\sum_{A=5}^6[c_1^O(K)]_{AA}\frac{1}{4\pi s\Gamma(s)}\right\}\qquad ,
\]
but the first Seeley coefficients due to bosonic and ghost
fluctuations, respectively, give:
\begin{eqnarray*}
\sum_{A=1}^4[c_1^I(K)]_{AA}&=&\int \, d^2x \,
[5-5|S_1|^2-3|S_2|^2-2V_kV_k]\\
\sum_{A=5}^6[c_1^O(K)]_{AA}&=&\int \, d^2x \,
[1-|S_1|^2-3|S_2|^2-2V_kV_k]\\
c_1(K^{\rm G})&=&\int \, d^2x \, [1-|S_1|^2-|S_2|^2] \qquad .
\end{eqnarray*}
Therefore,
\begin{eqnarray*}
\Delta M_{TS}^{(1)C} (-1/2) &=& - {\hbar m \over 16 \pi }\left\{
\left(\Sigma^{(1)} (S,V_k)+2\int \, d^2x \, \left|S_2\right|^2
\right) \cdot {\gamma[-1/2,1] \over \Gamma(1/2)}\right. \\
&-&\left.\left(\Sigma^{(0)} (S,V_k)-2\int \, d^2x \,
\left|S_2\right|^2\right)\cdot \frac{2}{\Gamma(1/2)} \right\} \qquad
.
\end{eqnarray*}
On the other hand, the contribution to the one-loop semi-local
string tension shift of the mass renormalization counter-terms
is{\footnote{$\gamma[-1/2,1]$ instead of $\Gamma[-1/2]$ is used in
this formula to be consistent with the approximation in $\Delta
M_{TS}^{(1)C} (-{1\over 2})$. }}:
\[
\Delta M_{TS}^R (1/2)  =  {\hbar m \over 16 \pi} \cdot\left\{
\Sigma^{(1)} (S,V_k) \cdot
 {\gamma[-1/2,1]\over
 \Gamma(1/2)}-{2\over\Gamma(1/2)}\cdot\Sigma^{(0)}(S,V_k)\right\}\qquad ,
\]
which almost cancels the contribution to the semi-local string
Casimir energy density of the $c_1$ coefficients. We finally obtain
the high-temperature one-loop semi-local topological soliton mass
shift formula:
\begin{eqnarray}
\Delta M_{TS}&=& -{\hbar m \over 16\pi\sqrt{\pi}} \left[
\sum_{n=2}^{N_0}\,\left\{ [\sum_{A=1}^4[c_n(K)]_{AA}-c_n(K^{\rm G})]
\cdot\gamma[n-\frac{3}{2},1]+\sum_{A=5}^6\frac{[c_n(K)]_{AA}}{n-\frac{3}{2}}\right\}+4
l\cdot 8\pi \right] \nonumber\\&-&\frac{\hbar
m}{8\pi\sqrt{\pi}}\cdot \int \, d^2x \,
\left|S_2\right|^2(x_1,x_2)\cdot \left(\gamma[-{1\over
2},1]-2\right)\quad . \label{eq:olslms}
\end{eqnarray}
In this final formula (\ref{eq:olslms}) there are four types of
terms:
\begin{itemize}
\item First, polynomial expressions in incomplete Gamma
functions times the heat-kernel expansion coefficients for K -
saving only the four first diagonal contributions due to massive
bosonic particles- and $K^G$ -coming from fermionic massive
particles. All of them start from the second-order coefficients.
\item Second, polynomial expressions in ${1\over n-{3\over 2}}$, including
the last two diagonal heat kernel coefficients that collect the
contribution of massless Goldstone particles. The starting
coefficients are also of second order.
\item Third, a factor
proportional to $4l$, taking into account the subtraction of the
$4l$ zero modes.
\item An extra piece proportional to the norm of the second
Higgs field due to the imperfect cancelation of the contribution of
first-order Seeley coefficients by mass renormalization
counter-terms.
\end{itemize}

Finally, let us mention that by cutting the expansion at a finite
number, $N_0$, we admit an error - besides the rejected entire parts
- which is a priori proportional to $\gamma[N_0-{1\over 2},1]\simeq
{1\over N_0-{1\over 2}}$, for $N_0$ large. Nevertheless, as we shall
see in the next Section, once all the calculations have been done
the degree of convergence of the results is very good, meaning that
the proportionality coefficient in that error is very small. The
reliability of the method is therefore quite high.

\section{Mathematica calculations}

\subsection*{Seeley densities for spherically symmetric semi-local vortices}

We shall apply (\ref{eq:olslms}) to spherically symmetric vortices.
The heat kernel local coefficients, however, depend on successive
derivatives of the solution. This dependence can increase the error
in the estimation of these local coefficients because we are
handling an interpolating polynomial as the numerically generated
solution, and the successive derivation with respect to $r$ of such
a polynomial introduces inaccuracies. Indeed, this operation is
plugged into the algorithm that generates the local coefficients in
order to speed up this process. It is thus of crucial importance to
use the first-order differential equations
(\ref{eq:sem1})-(\ref{eq:sem2})-(\ref{eq:sem3}) in order to
eliminate the derivatives of the solution and to write the local
coefficients as expressions that depend only on the fields.
Recalling the form of the spherically symmetric solutions,
\begin{eqnarray*}
S_1^1(x_1,x_2)&=&f(r)\cos\theta \hspace{2cm}S_1^2(x_1,x_2)=f(r)\sin\theta \\
S_2^1(x_1,x_2)&=&h(r)\hspace{2.9cm}S_2^2(x_1,x_2)=0 \\
V_1(x_1,x_2)&=&-\frac{\alpha(r)}{r}\sin\theta \hspace{1.7cm}
V_2(x_1,x_2)=\frac{\alpha(r)}{r}\cos\theta \qquad ,
\end{eqnarray*}
use of the first-order equations shows that:
{\small\begin{eqnarray*} \frac{\partial S_1^1}{\partial
x_1}&=&\frac{
f(r)}{r}\left[1-\alpha(r)\cos^2\theta\right]\hspace{1cm}
\frac{\partial S_1^2}{\partial
x_2}=\frac{f(r)}{r}\left[1-\alpha(r)\sin^2\theta
\right]\hspace{1cm}\frac{\partial S_2^1}{\partial x_2}=-\frac{
h(r)}{r}\alpha(r)
\cos\theta \\
\frac{\partial S_1^1}{\partial x_2}&=&-\frac{ f(r)}{r}\alpha(r)
\cos\theta \sin \theta\hspace{1.2cm}\frac{\partial S_1^2}{\partial
x_1}=-\frac{ f(r)}{r}\alpha(r)) \sin\theta \cos
\theta\hspace{1cm}\frac{\partial S_2^1}{\partial x_1}=-\frac{
h(r)}{r}\alpha(r)) \sin\theta
\\\frac{\partial V_1}{\partial x_1}&=&{\cos
2\theta\over 2}\left[\frac{2 f(r)\alpha(r)}{r}+{f^2(r)+h^2(r)-1\over
2}\right] \hspace{0.4cm} \quad \frac{\partial V_1}{\partial x_2}=-
\cos
2\theta\frac{\alpha(r)}{r^2}+{\sin^2\theta\over 2}(f^2(r)+h^2(r)-1)\\
\frac{\partial V_2}{\partial x_1}&=&- \cos
2\theta\frac{\alpha(r)}{r^2}- {\cos^2\theta\over 2}
(f^2(r)+h^2(r)-1)\hspace{0.3cm}\hspace{0.5cm} \frac{\partial
V_2}{\partial x_2}=-{\cos 2\theta\over 2}\left[\frac{2
f(r)\alpha(r)}{r}+{f^2(r)+h^2(r)-1\over 2}\right] \qquad .
\end{eqnarray*}}
Bearing this in mind, we solve the recurrence relations to find:

{\scriptsize \begin{eqnarray*}
{\rm tr}\,c_1^I(r) &=&5 - \frac{2\,{\alpha(r)}^2}{r^2} - 5\,{f(r)}^2 - 3\,{h(r)}^2 \\
{\rm tr}\,c_2^I(r)&=&\frac{1}{12\,r^4} \left[4\,{\alpha(r)}^4 +
27\,r^4\,{f(r)}^4 -
    8\,r^2\,\alpha(r)\,\left( -1 + 14\,{f(r)}^2 + {h(r)}^2 \right)  + \right. \\
    & & \hspace{0.8cm} +
    8\,{\alpha(r)}^2\,\left( -2 - 3\,r^2 + 9\,r^2\,{f(r)}^2 + 3\,r^2\,{h(r)}^2 \right)
    + \\ & & \hspace{0.8cm} + \left.
    {f(r)}^2\,\left( 56\,r^2 - 64\,r^4 + 34\,r^4\,{h(r)}^2 \right)  +
    r^4\,\left( 37 - 32\,{h(r)}^2 + 7\,{h(r)}^4 \right) \right] \\
{\rm tr}\,c_3^I(r) &=& \frac{1}{120\,r^6} \left\{-4\,{\alpha(r)}^6 -
4\,r^2\,{\alpha(r)}^3\,\left( 14 + 35\,{f(r)}^2 - 36\,{h(r)}^2
\right) + \right.
\\ && \hspace{0.9cm} +
    4\,{\alpha(r)}^4\,\left( 20 + 9\,r^2 + 32\,r^2\,{f(r)}^2 + 26\,r^2\,{h(r)}^2 \right)
     - \\ && \hspace{0.9cm} -
    2\,r^2\,\alpha(r)\,\left[ 57\,r^2\,{f(r)}^4 +
       {f(r)}^2\,\left( 32 + 331\,r^2 - 75\,r^2\,{h(r)}^2 \right)  -
       4\,\left( -1 + {h(r)}^2 \right) \,\left( -16 - 9\,r^2 + r^2\,{h(r)}^2 \right)
       \right]
       + \\ && \hspace{0.9cm} +
     {\alpha(r)}^2\,\left[ -256 - 144\,r^2 - 117\,r^4 + 99\,r^4\,{f(r)}^4 - 16\,r^2\,{h(r)}^2 +
       94\,r^4\,{h(r)}^2 - 61\,r^4\,{h(r)}^4 + \right. \\ && \hspace{2.3cm}
       + \left.
       2\,r^2\,{f(r)}^2\,\left( 56 + 183\,r^2 + 19\,r^2\,{h(r)}^2 \right)  \right]
       + \\ && \hspace{0.9cm} +
    r^4\,\left[ -16 + 151\,r^2 - 29\,r^2\,{f(r)}^6 + \left( 32 - 135\,r^2 \right) \,{h(r)}^2 +
       \left( -16 + 23\,r^2 \right) \,{h(r)}^4 + r^2\,{h(r)}^6 + \right. \\
       &&\hspace{1.2cm} + \left. \left.
       {f(r)}^4\,\left( -20 + 199\,r^2 - 57\,r^2\,{h(r)}^2 \right)  +
       {f(r)}^2\,\left( 392 - 321\,r^2 + 2\,\left( -68 + 111\,r^2 \right) \,{h(r)}^2 -
          27\,r^2\,{h(r)}^4 \right)  \right] \right\}
\end{eqnarray*}}

{\scriptsize\begin{eqnarray*} {\rm tr}\,c_1^O(r) &=& 1 -
\frac{2\,{\alpha(r)}^2}{r^2}
- {f(r)}^2 - 3\,{h(r)}^2 \\
{\rm tr}\,c_2^O(r) &=& \frac{1}{12\,r^4} [4\,{\alpha(r)}^4 -
r^4\,{f(r)}^4 + 8\,r^2\,\alpha(r)\,\left( 1 + 2\,{f(r)}^2 - {h(r)}^2
\right)  - \\ && \hspace{0.8cm} -
    8\,{\alpha(r)}^2\,\left( 2 + r^2 + r^2\,{f(r)}^2 - 5\,r^2\,{h(r)}^2 \right)  +
    2\,r^2\,{f(r)}^2\,\left( -4 + 9\,r^2\,{h(r)}^2 \right)  + \\ &&
    \hspace{0.8cm} +
    r^4\,\left( 1 - 8\,{h(r)}^2 + 19\,{h(r)}^4 \right) ] \\
{\rm tr}\,c_3^O(r) &=& \frac{-1}{120\,r^6}\left\{4\,{\alpha(r)}^6 -
4\,r^2\,{\alpha(r)}^3\,\left( -14 + 9\,{f(r)}^2 + 84\,{h(r)}^2
\right) - \right.
\\ & & \hspace{0.8cm} -
    4\,{\alpha(r)}^4\,\left( 20 + 3\,r^2 + 2\,r^2\,\left( {f(r)}^2 + 4\,{h(r)}^2 \right)  \right)
    + \\ && \hspace{0.8cm} +
    {\alpha(r)}^2\,\left[ 256 + 48\,r^2 - 3\,r^4 + 45\,r^4\,{f(r)}^4 +
       2\,r^2\,\left( -40 + 89\,r^2 \right) \,{h(r)}^2 - 115\,r^4\,{h(r)}^4
       + \right. \\ && \hspace{2.5cm} + \left.
       2\,r^2\,{f(r)}^2\,\left( 8 + 5\,r^2 - 35\,r^2\,{h(r)}^2 \right)
       \right]
       - \\ && \hspace{0.8cm} -
    2\,r^2\,\alpha(r)\,\left[ 53\,r^2\,{f(r)}^4 +
       4\,\left( -1 + {h(r)}^2 \right) \,\left( -16 - 3\,r^2 + 7\,r^2\,{h(r)}^2 \right)  -
       {f(r)}^2\,\left( 32 + 17\,r^2 + 47\,r^2\,{h(r)}^2 \right)
       \right]
       + \\ & & \hspace{0.8cm} +
    r^4\,\left[ 16 + 3\,r^2 + 3\,r^2\,{f(r)}^6 - \left( 32 + 19\,r^2 \right) \,{h(r)}^2 +
       \left( 16 + 23\,r^2 \right) \,{h(r)}^4 + 33\,r^2\,{h(r)}^6 +
       \right.
       \\ && \hspace{1.8cm} + \left. \left.
       {f(r)}^4\,\left( 52 - r^2 + 39\,r^2\,{h(r)}^2 \right)  +
       {f(r)}^2\,\left( -24 - 5\,r^2 + \left( -72 + 22\,r^2 \right) \,{h(r)}^2 +
          69\,r^2\,{h(r)}^4 \right)  \right] \right\}
\end{eqnarray*}}

{\scriptsize\begin{eqnarray*}
c_1^G (r) &=& 1 - {f(r)}^2 - {h(r)}^2 \\
c_2^G(r) &=& \frac{1}{6\,r^2} \left[3\,r^2 + 2\,r^2\,{f(r)}^4 -
\left( 5\,r^2 + 4\,{\alpha(r)}^2 \right) \,{h(r)}^2 +
2\,r^2\,{h(r)}^4 + \right. \\ & & \hspace{0.8cm} \left. +
{f(r)}^2\,\left( -4 - 5\,r^2 + 8\,\alpha(r) - 4\, {\alpha(r)}^2 +
4\,r^2\,{h(r)}^2 \right) \right] \\
c_3^G(r) &=& \frac{-1}{60\,r^4}\left\{-10\,r^4 + 4\,r^4\,{f(r)}^6 +
\left[ 23\,r^4 - 8\,r^2\,\alpha(r) + 16\,\left( 1 + r^2 \right)
\,{\alpha(r)}^2 + 32\,{\alpha(r)}^3 + 16\,{\alpha(r)}^4 \right]
\,{h(r)}^2 + \right.
\\ & &  \hspace{0.9cm} + r^2\,\left[ -17\,r^2 + 8\,\alpha(r) - 16\,{\alpha(r)}^2 \right] \,{h(r)}^4 +
4\,r^4\,{h(r)}^6 +
 \\  & &  \hspace{0.9cm} +    r^2\,{f(r)}^4\,\left[ -24 - 17\,r^2 + 40\,\alpha(r) -
 16\,{\alpha(r)}^2 + 12\,r^2\,{h(r)}^2 \right]
 +  \\ &&  \hspace{0.9cm} + {f(r)}^2\,\left[ -32\,{\alpha(r)}^3 + 16\,{\alpha(r)}^4 +
 8\,r^2\,\alpha(r)\,\left( -5 + 6\,{h(r)}^2 \right)  +
 16\,{\alpha(r)}^2\,\left( 1 + r^2 - 2\,r^2\,{h(r)}^2 \right) \right. + \\
& & \left. \hspace{0.9cm} + \left. r^2\,\left( 24 + 23\,r^2 -
2\,\left( 10 + 17\,r^2 \right) \,{h(r)}^2 + 12\,r^2\,{h(r)}^4
\right)  \right] \right\} \qquad ,
\end{eqnarray*}}
etcetera, by means of a computing program implemented on a PC with
Mathematica.

\subsubsection*{One-loop mass shift for the mass of the ANO vortex}

Denoting simply
\[
{\rm tr}\,c_n^I=2\pi \int_0^\infty \, dr \, r {\rm
tr}\,c_n^I(r)\quad , \quad {\rm tr}\,c_n^O=2\pi \int_0^\infty \, dr
\, r {\rm tr}\,c_n^O(r)\quad , \quad  c_n^G=2\pi \int_0^\infty \, dr
\, r c_n^G(r) \qquad ,
\]
by plugging the NO vortex solution in these expressions, i.e. the
case $h_0=0.0$, embedded in this model, we find the Table at the
left:

\begin{center}
\begin{tabular}{|c|c|c|c|}
 \hline  & \multicolumn{3}{|c|}{$h_0=0.0$}  \\ \hline
 $n$ & ${\rm tr}\,c_n^I$ & ${\rm tr}\,c_n^O$ & $c_n^G$  \\ \hline
 1 & -41.4469  & -91.8429   & 12.599 \\ \hline
 2 & 30.3736  & 0.96286     & 2.61518  \\ \hline
 3 & 12.9447  & -0.0592415  & 0.32005 \\ \hline
 4 & 4.22603  & 0.001512548  & 0.0230445 \\ \hline
 5 & 1.05059  & 0.000758663 & 0.0013023 \\ \hline
 6 & 0.20900  & -0.00023912 & 0.0000698185 \\ \hline
\end{tabular}
\hspace{1cm}
\begin{tabular}{|c|c|}  \hline
$N_0$ & $\Delta M_V (N_0)$ \\
 & $l=1$
\\ \hline
2 & -1.61536  \\
3 & -1.66862 \\
4 & -1.67809 \\
5 & -1.67966 \\
6 & -1.67989 \\ \hline
\end{tabular}
\vspace*{0.2cm}

\textit{(left) The sixth lowest Seeley coefficients for
Nielsen-Olesen self-dual vortices. (right) Convergence of the
one-loop mass shift for self-dual semi-local NO vortices in units of
$\hbar m$.}
\end{center}
Because the NO vortex solutions have been generated numerically,
integration over the whole plane of the Seeley densities can also
only be performed numerically. Therefore, we are forced to put a
cut-off into the area and replace the infinite plane by a discus of
radius $R$, which in the calculations above was chosen to be
$R=10.000$ (compare with the profiles of Figure 1). Use of these
numbers in formula (\ref{eq:olslms}) provides us with the Table at
the right, where the one-loop mass shifts in $\hbar m$ units of
semi-local self-dual NO vortices are shown up to sixth order in the
asymptotic formula.

Our result for the one-loop mass shift of semi-local self-dual NO
vortices is:
\begin{equation}
\Delta M_V (N_0=6)= -1.67989 \hbar m \qquad \qquad .
\label{eq:qcslsdv}
\end{equation}
The ratio between the mass shifts of self-dual NO vortices in the
semi-local and normal Abelian Higgs model is:
\[
\frac{\Delta M_V^{\rm SLAHM}}{\Delta M_V^{\rm
AHM}}=\frac{1.67989}{1.09449}=1.53486 \qquad ,
\]
see \cite{AGMT}. Similar relations exist between ratios of kink mass
shifts in the $\lambda(\phi)^4_2$ model and the BNRT model; a
two-field system that depends on a real positive parameter $\sigma$:
\[
\frac{\Delta M_K^{\rm BNRTM}}{\Delta M_K^{\rm \Lambda\Phi
M}}=\frac{0.693943}{0.471113}=1.47299 \, \, \quad {\rm if} \quad
\sigma=0.99
\]
\[
\frac{\Delta M_K^{\rm BNRTM}}{\Delta M_K^{\rm \Lambda\Phi
M}}=\frac{0.698445}{0.471113}=1.48254 \, \, \quad {\rm if} \quad
\sigma=1.01 \qquad ,
\]
see \cite{Aai3}.

Several comments are in order:
\begin{enumerate}

\item The coefficients ${\rm tr}\, c_n^I$, for $n=2,3, \cdots, 6$,
are identical, within numerical precision, to the same coefficients
for self-dual NO vortices in the Abelian Higgs model. Therefore, the
difference of the vortex mass shift arising in the semi-local
Abelian Higgs model is due to the contribution of the ${\rm tr}\,
c_n^O$ coefficients and to the double number of zero modes.

\item ${\rm tr}\, c_1^I$ and ${\rm tr}\, c_1^O$ are very large
negative numbers. In the case at hand, $h_0=0$, their contribution
cancels against the energy induced by mass renormalization
counter-terms. A similar behavior of $h_0\neq 0$ self-dual
semi-local topological solitons would mean that rather than
decreasing their mass should grow as the result of one-loop
fluctuations, a possibility that we shall study below.

\end{enumerate}

\subsection*{Seeley coefficients for semi-local topological
solitons:\\ one quantum of magnetic flux and various values of
$h_0$.}

To accomplish this goal, we offer several Tables in the Appendix
with similar numerical calculations on discuses of radius $R=10^2$,
$R=10^3$, $R=10^4$, $R=10^5$, and $R=10^6$ for the values $h_0=0$,
$h_0=0.1$, $h_0=0.3$, $h_0=0.6$, and $h_0=0.9$. Inspection of these
Tables raises some points that merit comment:
\begin{enumerate}
\item We see from the Tables for the $h_0=0.0$ case that although the
first-order coefficients grow spectacularly with the radius, the
highest coefficients are quite stable against growing areas of
integration, suggesting good behavior -a finite value very close to
the value obtained at $R=10^4$- of the one-loop correction in the
infinite area limit.
\item Things start to be different when we consider the $h_0=0.1$ case.
First, the positive and negative contributions of the first-order
Seeley coefficients are not completely canceled out by mass
renormalization counter-terms. Second, of special importance is the
departure of the values of ${\rm tr}c_n^O(K)$ from the same numbers
for the NO vortex, whereas ${\rm tr}c_n^I(K)$ and $c_n(K^G)$ change
slowly with $h_0$. Moreover, ${\rm tr}c_1^O(K)$ for $h_0=0.1$,
besides not  being exactly canceled, is negative and large. This
effect prompts a increasingly less negative one-loop correction,
eventually changing sign, with larger and larger $R$. This behavior
differs completely from that of embedded NO vortices and suggests a
different fate for semi-local topological solitons when quantum
fluctuations enter the game.
\item The numbers offered in the Tables for $h_0=0.3$ follow a similar pattern to
those of the $h_0=0.1$ semi-local topological soliton. There are,
however, some quantitative differences. ${\rm tr}c_n^I(K)$, $n\neq
1$, $h_0=0.3$, differs from the same number of the NO vortex more
than the $h_0=0.1$ topological soliton. Thus, these numbers grow
with $h_0$ and $R$. ${\rm tr}c_1^I(K)$ decreases with respect to the
same number for the NO vortex and with $R$. There is a change of
sign at $R=10^3$ if $h_0=0.1$ and at $R=10^4$ if $h_0=0.3$. The
numbers ${\rm tr}c_1^O(K)$ behave as in the $h_0=0.1$ case but
depart more from NO values. If $h_0=0.1$ the ghost coefficients
$c_n(K^G)$ are slightly larger than these numbers for $h_0=0$, but
they start to decrease (rapidly with $R$) when $h_0=0.3$.
\item The pattern observed for the evolution of the Seeley coefficients
when the area grows in the cases $h_0=0.1$ and $h_0=0.3$ is
reinforced for broader semilocal topological defects with $h_0=0.6$
and $h_0=0.9$. The unbalanced first-order coefficients ${\rm tr}
c_1^O$, obtained by by mass renormalizations, rapidly tend toward
huge negative values.
\end{enumerate}

\subsubsection*{Mass shift of semi-local topological solitons}

The consequence is as follows: whereas the one-loop mass shift of
embedded ANO vortices is always negative and varies extremely slowly
as the area increases towards more negative values, one-loop mass
shifts of genuine semilocal topological solitons with $|h_0|>0$
become less negative, and even positive, for larger areas, as is
shown in the following Table.\vspace{0.2cm}

{\footnotesize\begin{center} \textit{One-loop mass shifts for
semi-local topological solitons: Five values of $h_0$, five values
of $R$, and fixed $N_0=6$.} \\ \vspace{0.5cm}
\begin{tabular}{|c|c|c|c|c|c|}  \hline
$R$ & $\Delta M_V (N_0=6,R)$ & $\Delta M_V (N_0=6,R)$ & $\Delta M_V
(N_0=6,R)$
& $\Delta M_V (N_0=6,R)$ &$\Delta M_V (N_0=6,R)$ \\
 & $h_0=0.0$ & $h_0=0.1$ & $h_0=0.3$ & $h_0=0.6$ & $h_0=0.9$
\\ \hline
$10^2$ & -1.67955 & -1.61672 & -1.05000 & 2.10142 & 24.6066 \\
$10^3$ & -1.67971& -1.58311 & -0.626167 &4.5485 &  42.7747 \\
$10^4$ & -1.67989 & -1.55133 & -0.252586 &6.41655 & 60.9433 \\
$10^5$ & -1.68005 & -1.51957 &  0.12086  &8.5741  & 79.1116 \\
$10^6$ & -1.68026 & -1.48779 &  0.49433  &10.7203 & 97.2798 \\
\hline
\end{tabular}
\end{center}}

The classical degeneracy in energy between semi-local topological
defects seems to be broken by one-loop fluctuations, the embedded
ANO vortices becoming the ground states in the topological sector of
one quantum of magnetic flux. It is remarkable how strong this
effect becomes for large $|h_0|$.

These numerical results find support in the following qualitative
arguments based in the analysis of the potentials arising in the
matrix Schrodinger operators governing small fluctuations around
semi-local self-dual topological solitons. The more pertinent,
diagonal, operators in $K$ and $K^G$ are:
\[
A=K^G=-\bigtriangleup+f^2(r)+h^2(r)=-\bigtriangleup+A(r)
\]
\[
B=-\bigtriangleup+{1\over 2}(3f^2(r)+h^2(r)+2{\alpha^2(r)\over
r^2}-1)=-\bigtriangleup+B(r)
\]
\[
C=-\bigtriangleup+{1\over 2}(f^2(r)+3h^2(r)+2{\alpha^2(r)\over
r^2}-1)=-\bigtriangleup+C(r) \quad .
\]
A glance at Figure 2 reveals:
\begin{enumerate}

\item For ANO vortices, $h_0=0.0$ and $h(r)=0$, there are attractive
potential wells with the bottom at the origin. Thus, the correction
must be negative. Also, the field profiles reach their vacuum values
exponentially and the corrections are quite stable with respect to
the area of the normalization box.

\item For $h_0=0.1$, $h(r)$ is not zero, see also Figure 1 and take
into account that: ${\alpha^2(r)\over r^2}\simeq_{r\rightarrow 0}
r^2$. We observe that the corresponding terms push the wells upwards
from the bottom, making them less attractive. This explains why the
mass shifts are less negative.

\item For $h_0=0.3$, $h(r)$ is big enough to globally produce a
change from attractive to repulsive potential forces provided that
the area of the normalization box is sufficiently large. There is
consequently a change in the sign of the mass correction.

\item For $h_0=0.6$ and $h_0=0.9$, potential barriers dominate
starting from relatively small areas of the normalization box. The
field profiles go to their vacuum values very slowly  and the
barriers become very wide, explaining the strong dependence on the
size of the normalization box.
\end{enumerate}

\begin{figure}[htdp]
\centerline{\includegraphics[height=3cm]{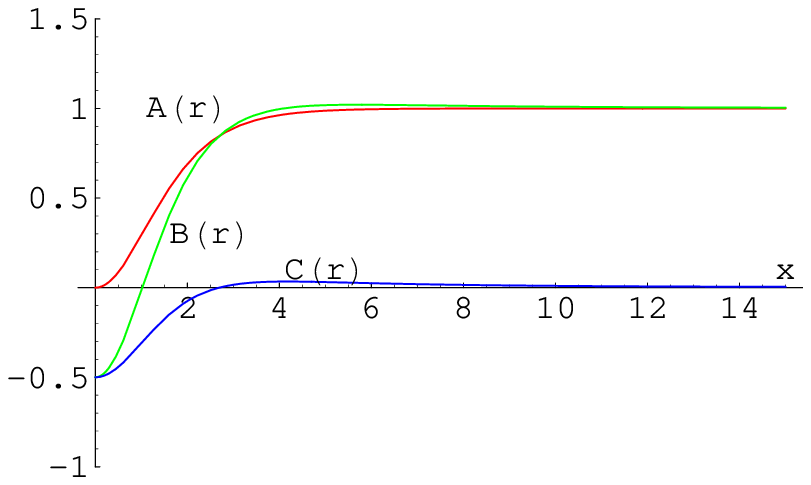}\hspace{0.3cm}
\includegraphics[height=3cm]{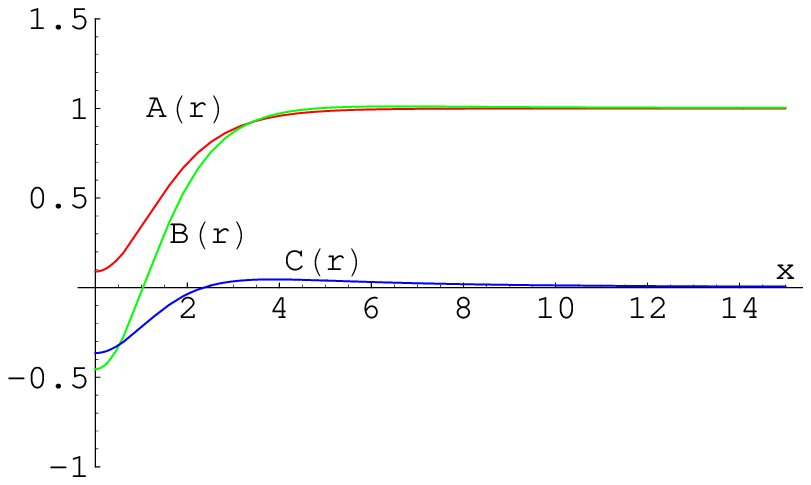}\hspace{0.3cm}
\includegraphics[height=3cm]{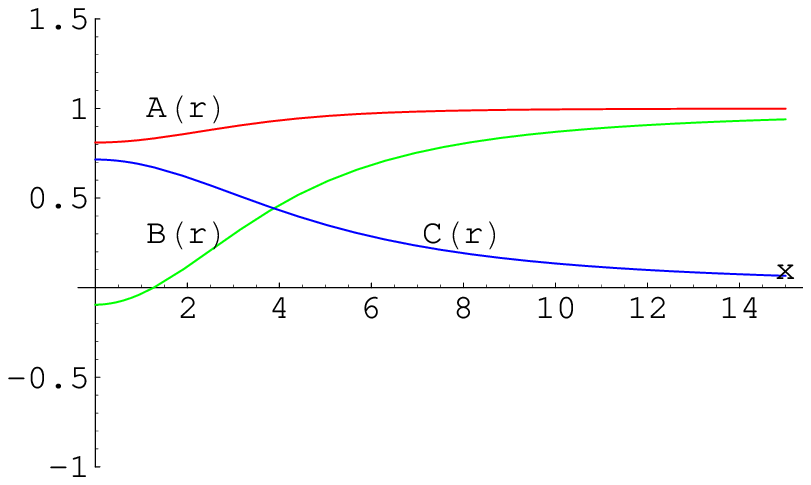}}
\centerline{\small \textit{a) Functions $A(r)$, $B(r)$ and $C(r)$
for a) $h_0=0.0$, a) $h_0=0.3$ and a) $h_0=0.9$}}
\end{figure}

We conclude that the classical degeneracy is broken by one-loop
fluctuations. Even if the classical topological bound is saturated
by all the solitons in the moduli space of solutions of the
first-order equations, quantum effects can distinguish between the
different energy densities of these extended objects. The two
extremes are the ANO vortices, $h_0=0$, where the energy density is
concentrated around the zero of the Higgs field, and the ${\rm
CP}^1$-lumps, with energy densities uniformly distributed over the
whole spatial plane. A similar effect has been observed before in
the moduli space of degenerated two-component kinks analyzed in
Reference \cite{Aai3}.

\subsubsection*{Infrared divergences: quantum fate of semi-local
topological solitons}

The dependence on the area of the normalization box is due to the
slow decay (non-exponential) to their vacuum values of genuine
semi-local topological solitons as compared with ANO vortices.
Plugging the asymptotic form of the spherically symmetric
topological soliton solutions obtained in Section \S. 1.7 into the
Seeley densities, we find the following behavior at infinity in
terms of the parameter $|h^1|$ (which sets the large $r$ behavior of
the solutions): {\small\begin{eqnarray*} 2\pi r{\rm
tr}c_1^I(r)&\simeq&{\atop r\rightarrow\infty }
-\frac{4\pi}{r}(1-|h^1|^2)+\frac{4\pi}{r^3}(12|h^1|^2-|h^1|^4)+{\cal
O}({1\over r^5}) \\ 2\pi r{\rm tr}c_1^O(r)&\simeq&{\atop
r\rightarrow\infty }
-\frac{4\pi}{r}(1+|h^1|^2)+\frac{4\pi}{r^3}(4|h^1|^2+|h^1|^4)+{\cal
O}({1\over r^5}) \\ 2\pi r c_1^G(r)&\simeq&{\atop r\rightarrow\infty
} \,\, \frac{8\pi}{r^3}|h^1|^2+{\cal O}({1\over r^5})
\end{eqnarray*}}
{\small\begin{eqnarray*} 2\pi r{\rm tr}c_2^I(r)&\simeq&{\atop
r\rightarrow\infty }
\,\,\frac{2\pi}{r}|h^1|^2+\frac{2\pi}{r^3}(-1+4|h^1|^2-|h^1|^4)+{\cal
O}({1\over r^5}) \\ 2\pi r{\rm tr}c_2^O(r)&\simeq&{\atop
r\rightarrow\infty } \,\,
\frac{2\pi}{r}|h^1|^2+\frac{2\pi}{r^3}(-1+4|h^1|^2-|h^1|^4)+{\cal
O}({1\over r^5}) \\2\pi r c_2^G(r)&\simeq&{\atop r\rightarrow\infty
} \,\,
\frac{64\pi}{3r^5}|h^1|^2+\frac{\pi}{r^7}(768|h^1|^2-80|h^1|^4)+{\cal
O}({1\over r^9})
\end{eqnarray*}}
{\small\begin{eqnarray*} 2\pi r{\rm tr}c_3^I(r)&\simeq&{\atop
r\rightarrow\infty }
\,\,\frac{2\pi}{3r}|h^1|^2+\frac{2\pi}{3r^3}(3|h^1|^2-|h^1|^4)+{\cal
O}({1\over r^5}) \\ 2\pi r{\rm tr}c_3^O(r)&\simeq&{\atop
r\rightarrow\infty } \,\,
-\frac{2\pi^2}{3r}|h^1|^2+\frac{2\pi}{3r^3}(-4|h^1|^2+|h^1|^4)+{\cal
O}({1\over r^5}) \\2\pi r c_3^G(r)&\simeq&{\atop r\rightarrow\infty
} \,\, \frac{384\pi}{5r^7}|h^1|^2+{\cal O}({1\over r^8})
\end{eqnarray*}}
{\small\begin{eqnarray*} 2\pi r{\rm tr}c_4^I(r)&\simeq&{\atop
r\rightarrow\infty }
\,\,\frac{\pi}{6r}|h^1|^2+\frac{\pi}{6r^3}({12\over
5}|h^1|^2-|h^1|^4)+{\cal O}({1\over r^5}) \\ 2\pi r{\rm
tr}c_4^O(r)&\simeq&{\atop r\rightarrow\infty } \,\,
-\frac{\pi}{6r}|h^1|^2+\frac{\pi}{6r^3}(4|h^1|^2-|h^1|^4)+{\cal
O}({1\over r^5}) \\2\pi r c_4^G(r)&\simeq&{\atop r\rightarrow\infty
} \,\,
\frac{832\pi}{21r^9}|h^1|^4-\frac{43\pi}{30r^9}|h^1|^6+\frac{\pi}{48r^9}|h^1|^8+
\frac{\pi}{768r^9}|h^1|^{10}+{\cal O}({1\over r^{11}})
\end{eqnarray*}}
{\small\begin{eqnarray*} 2\pi r{\rm tr}c_5^I(r)&\simeq&{\atop
r\rightarrow\infty }
\,\,\frac{\pi}{30r}|h^1|^2+\frac{\pi}{30r^3}(2|h^1|^2-|h^1|^4)+{\cal
O}({1\over r^5}) \\ 2\pi r{\rm tr}c_5^O(r)&\simeq&{\atop
r\rightarrow\infty } \,\,
-\frac{\pi}{30r}|h^1|^2+\frac{\pi}{30r^3}(-4|h^1|^2+|h^1|^4)+{\cal
O}({1\over r^5})
\\2\pi r c_5^G(r)&\simeq&{\atop r\rightarrow\infty } \,\,-\frac{\pi}{r^9}
\left(\frac{2048}{105}|h^1|^2-\frac{416}{189}|h^1|^4+\frac{43}{540}|h^1|^6-{1\over
864}|h^1|^8+{1\over 138240}|h^1|^{10}\right)+{\cal O}({1\over
r^{11}})
\end{eqnarray*}}
{\small\begin{eqnarray*} 2\pi r{\rm tr}c_6^I(r)&\simeq&{\atop
r\rightarrow\infty } \,\,\frac{\pi}{180 r}|h^1|^2+\frac{\pi}{15
r^3}({1\over 7}|h^1|^2-{1\over 12}|h^1|^4)+{\cal O}({1\over r^5})
\\ 2\pi r{\rm tr}c_6^O(r)&\simeq&{\atop r\rightarrow\infty } \,\,
\frac{\pi}{180r}|h^1|^2+\frac{\pi}{45r^3}(|h^1|^2-{1\over
4}|h^1|^4)+{\cal O}({1\over r^5})
\\
\end{eqnarray*}}
The key observation is the appearance of infrared logarithmic
divergences in the Seeley coefficients ${\rm tr}c^I_n(K)$ and ${\rm
tr}c^O_n(K)$ for all $n$. The ghost coefficients $c^G_n(K^G)$,
however, are infrared convergent. The combination of the signs that
we have seen in the previous sub-Sections and the large $r$ behavior
show that one-loop mass shifts of semi-local topological solitons
tend to $+\infty$ at the infinite area limit. Semi-local topological
defects grow infinitely massive due to the infrared effects of
one-loop fluctuations. This phenomenon seems to be amazingly close
to the non-existence of Goldstone bosons in (1+1)-dimensions.

There is a very important exception: for ANO vortices, $|h^1|=0$ and
only the first-order coefficients are infrared divergent. However,
the contribution of these coefficients is totally canceled by mass
renormalization counter-terms. Our results suggest that only the ANO
vortices between all the semi-local topological solitons survive
one-loop quantum fluctuations. It would be very interesting to try a
more analytic approach to this problem in order to fully elucidate
this delicate issue.
\section*{Appendix}
\subsubsection*{Semilocal strings with $h_0=0.0$. Embedded NO vortices}
{\small\begin{center}
\textit{Seeley coefficients: Semilocal strings with $h_0=0.0$} \\
\begin{tabular}{|c|c|c|c|}
 \hline  & \multicolumn{3}{|c|}{$h_0=0.0$\, , \, $R=10^2$}  \\ \hline
 $n$ & ${\rm tr}\,c_n^I$ & ${\rm tr}\,c_n^O$ & $c_n^G$  \\ \hline
 1 & 16.3087  & -33.9956   &  12.5761 \\ \hline
 2 & 30.3548  &   0.959353   & 2.61136  \\ \hline
 3 & 12.9428  &  -0.0596236  & 0.319668 \\ \hline
 4 &  4.2259   &  0.00149819 & 0.0230172 \\ \hline
 5 &  1.050558 &  0.000757146 &0.00122871 \\ \hline
 6 &  0.209003 & -0.000239188& 0.0000697495 \\ \hline
\end{tabular}
\begin{tabular}{|c|c|c|c|}
 \hline  & \multicolumn{3}{|c|}{$h_0=0.0$ \, , \, $R=10^3$}  \\ \hline
 $n$ & ${\rm tr}\,c_n^I$ & ${\rm tr}\,c_n^O$ & $c_n^G$  \\ \hline
 1 & -12.5691 & -62.9193 &     12.5875 \\ \hline
 2 &  30.3641 &  0.960952 &     2.61327   \\ \hline
 3 &  12.9437  & -0.0594326 &   0.319859 \\ \hline
 4 &   4.22596 &  0.00151183 &  0.0230309 \\ \hline
 5 &   1.05058  & 0.0000747904 &0.00122947 \\ \hline
 6 &   0.209003 & -0.000239154 & 0.000069784 \\ \hline
\end{tabular}
\begin{tabular}{|c|c|c|c|}
 \hline  & \multicolumn{3}{|c|}{$h_0=0.0$ \, , \, $R=10^4$}  \\ \hline
 $n$ & ${\rm tr}\,c_n^I$ & ${\rm tr}\,c_n^O$ & $c_n^G$  \\ \hline
 1 & -41.4469  & -91.8429   & 12.599 \\ \hline
 2 & 30.3736  & 0.96286     & 2.61518  \\ \hline
 3 & 12.9447  & -0.0592415  & 0.32005 \\ \hline
 4 & 4.22603  & 0.001512548  & 0.0230445 \\ \hline
 5 & 1.05059  & 0.000758663 & 0.0013023 \\ \hline
 6 & 0.20900  & -0.00023912 & 0.0000698185 \\ \hline
\end{tabular}
\begin{tabular}{|c|c|c|c|}
 \hline  & \multicolumn{3}{|c|}{$h_0=0.0$ \, , \, $R=10^5$}  \\ \hline
 $n$ & ${\rm tr}\,c_n^I$ & ${\rm tr}\,c_n^O$ & $c_n^G$  \\ \hline
 1 & -70.3247 & -120.767    & 12.6105 \\ \hline
 2 &  30.3829 &  0.964771    & 2.61709  \\ \hline
 3 &  12.9456 & -0.0590504   & 0.320242 \\ \hline
 4 &   4.2261 &  0.00153913  & 0.0210582 \\ \hline
 5 &   1.05059 & 0.000759469 & 0.00123099 \\ \hline
 6 &   0.209003&-0.000239086 & 0.00002698506 \\ \hline
\end{tabular}
\begin{tabular}{|c|c|c|c|}
 \hline  & \multicolumn{3}{|c|}{$h_0=0.0$ \, , \, $R=10^6$}  \\ \hline
 $n$ & ${\rm tr}\,c_n^I$ & ${\rm tr}\,c_n^O$ & $c_n^G$  \\ \hline
 1 & -99.2028 & -149.690     & 12.6220 \\ \hline
 2 &  30.3920 &    0.9666682  &  2.61901 \\ \hline
 3 &   12.9466 &  -0.0588602 &   0.320432 \\ \hline
 4 &    4.22617&   0.00155181&   0.02300754 \\ \hline
 5 &    1.05058 &  0.0007600532& 0.00123155 \\ \hline
 6 &    0.209003& -0.00023874  & 0.0000695093  \\ \hline
\end{tabular}
\end{center}}
\subsubsection*{Semilocal strings with $h_0=0.1$}
{\small\begin{center}
\textit{Seeley coefficients: Semilocal strings} \\
\begin{tabular}{|c|c|c|c|}
 \hline  & \multicolumn{3}{|c|}{$h_0=0.1$\, , \, $R=10^2$}  \\ \hline
 $n$ & ${\rm tr}\,c_n^I$ & ${\rm tr}\,c_n^O$ & $c_n^G$  \\ \hline
 1 & 19.0369  &  -35.1867  &  12.8061 \\ \hline
 2 & 31.2832  &    1.69082   & 2.62117  \\ \hline
 3 & 13.2000  &  -0.277417  & 0.31637 \\ \hline
 4 &  4.28364   &  0.567246 &   0.0225427 \\ \hline
 5 &  1.06171 &   -0.0102584 &  0.00119161 \\ \hline
 6 &  0.210813 &   0.00159792&  0.0000667653 \\ \hline
\end{tabular}
\begin{tabular}{|c|c|c|c|}
 \hline  & \multicolumn{3}{|c|}{$h_0=0.1$ \, , \, $R=10^3$}  \\ \hline
 $n$ & ${\rm tr}\,c_n^I$ & ${\rm tr}\,c_n^O$ & $c_n^G$  \\ \hline
 1 & -7.67553  &-64.6429   &13.0898 \\ \hline
 2 & 31.9215 &   2.14009    &  2.66844  \\ \hline
 3 & 13.3577  & -0.406805  &  0.321364 \\ \hline
 4 &  4.31885  & 0.090590  &  0.0228803 \\ \hline
 5 &  1.0685  & -0.01694523 & 0.00121037 \\ \hline
 6 &  0.211935  &0.00271637 & 0.000067618\\ \hline
\end{tabular}
\begin{tabular}{|c|c|c|c|}
 \hline  & \multicolumn{3}{|c|}{$h_0=0.1$ \, , \, $R=10^4$}  \\ \hline
 $n$ & ${\rm tr}\,c_n^I$ & ${\rm tr}\,c_n^O$ & $c_n^G$  \\ \hline
 1 & -34.3878  & -94.099 &     13.3734 \\ \hline
 2 &  32.5602 &   2.58966 &     2.71571   \\ \hline
 3 &  13.5154  & -0.536179 &    0.326091 \\ \hline
 4 &   4.35407 &  0.124453 &    0.023218 \\ \hline
 5 &   1.0753  & -0.0236316 &   0.00122913 \\ \hline
 6 &   0.213057 & 0.00383473   &0.0000684707 \\ \hline
\end{tabular}
\begin{tabular}{|c|c|c|c|}
 \hline  & \multicolumn{3}{|c|}{$h_0=0.1$ \, , \, $R=10^5$}  \\ \hline
 $n$ & ${\rm tr}\,c_n^I$ & ${\rm tr}\,c_n^O$ & $c_n^G$  \\ \hline
 1 & -61.1001 & -123.555    &13.6571 \\ \hline
 2 &  33.1989 &  3.03924    & 2.76299  \\ \hline
 3 &  13.6732 & -0.665552   & 0.330819 \\ \hline
 4 &   4.38928 & 0.158316  &  0.0235856 \\ \hline
 5 &   1.0821 & -0.0303191 &  0.00124789 \\ \hline
 6 &   0.214679& 0.0049531 &  0.0000693256 \\ \hline
\end{tabular}
\begin{tabular}{|c|c|c|c|}
 \hline  & \multicolumn{3}{|c|}{$h_0=0.1$ \, , \, $R=10^6$}  \\ \hline
 $n$ & ${\rm tr}\,c_n^I$ & ${\rm tr}\,c_n^O$ & $c_n^G$  \\ \hline
 1 & -87.8124 & -153.011  & 13.9407 \\ \hline
 2 &  33.8375 &    3.48881 & 2.81037 \\ \hline
 3 &  13.8312 &   -0.784926 & 0.335521 \\ \hline
 4 &   4.42449&    0.192179&  0.00238892 \\ \hline
 5 &   1.0889 &  -0.030047& 0.00126624\\ \hline
 6 &   0.215306&  0.00607261  & 0.0000689434  \\ \hline
\end{tabular}
\end{center}}
\subsubsection*{Semilocal strings with $h_0=0.3$}
{\small\begin{center}
\textit{Seeley coefficients: Semilocal strings} \\
\begin{tabular}{|c|c|c|c|}
 \hline  & \multicolumn{3}{|c|}{$h_0=0.3$\, , \, $R=10^2$}  \\ \hline
 $n$ & ${\rm tr}\,c_n^I$ & ${\rm tr}\,c_n^O$ & $c_n^G$  \\ \hline
 1 & 27.8909 &  -48.6659  &  11.7815 \\ \hline
 2 & 39.9546  &   7.69438 &   2.22326  \\ \hline
 3 & 15.2219  &  -2.2614  &   0.247514 \\ \hline
 4 &  4.77643   & 0.540926&   0.0159336 \\ \hline
 5 &  1.15832 &  -0.10716 &   0.000764659 \\ \hline
 6 &  0.226592 &  0.0176963&  0.0000395987 \\ \hline
\end{tabular}
\begin{tabular}{|c|c|c|c|}
 \hline  & \multicolumn{3}{|c|}{$h_0=0.3$ \, , \, $R=10^3$}  \\ \hline
 $n$ & ${\rm tr}\,c_n^I$ & ${\rm tr}\,c_n^O$ & $c_n^G$  \\ \hline
 1 &  2.4708  & -86.5124  &   10.8818 \\ \hline
 2 & 40.2110  &  11.5509    &  2.07325  \\ \hline
 3 & 16.4824  & -3.61196  &    0.232514 \\ \hline
 4 &  5.10495 &  0.873748  &   0.0148622 \\ \hline
 5 &  1.22480  &-0.173998 &    0.000705138 \\ \hline
 6 &  0.237707 & 0.0288234 &   0.0000368937\\ \hline
\end{tabular}
\begin{tabular}{|c|c|c|c|}
 \hline  & \multicolumn{3}{|c|}{$h_0=0.3$ \, , \, $R=10^4$}  \\ \hline
 $n$ & ${\rm tr}\,c_n^I$ & ${\rm tr}\,c_n^O$ & $c_n^G$  \\ \hline
 1 & -22.9489  & -124.360  &     9.98213 \\ \hline
 2 &  43.4678  &   15.4073 &     1.92331    \\ \hline
 3 &  17.7429  &   -4.96241 &    0.21752 \\ \hline
 4 &   5.43346 &    1.20654 &    0.0137912 \\ \hline
 5 &   1.29128  &  -0.24083 &    0.000645635 \\ \hline
 6 &   0.248822 &   0.0399494   &0.0000341891 \\ \hline
\end{tabular}
\begin{tabular}{|c|c|c|c|}
 \hline  & \multicolumn{3}{|c|}{$h_0=0.3$ \, , \, $R=10^5$}  \\ \hline
 $n$ & ${\rm tr}\,c_n^I$ & ${\rm tr}\,c_n^O$ & $c_n^G$  \\ \hline
 1 & -48.3715 & -162.207    & 9.08246 \\ \hline
 2 &  46.7241 &   19.2687   & 1.77336  \\ \hline
 3 &  19.0034 &  -6.31286   & 0.202525 \\ \hline
 4 &   5.76197 &  1.53933  &  0.0127201 \\ \hline
 5 &   1.35776 & -0.30766 &   0.000586148 \\ \hline
 6 &   0.259938&  0.0510755 & 0.0000314752 \\ \hline
\end{tabular}
\begin{tabular}{|c|c|c|c|}
 \hline  & \multicolumn{3}{|c|}{$h_0=0.3$ \, , \, $R=10^6$}  \\ \hline
 $n$ & ${\rm tr}\,c_n^I$ & ${\rm tr}\,c_n^O$ & $c_n^G$  \\ \hline
 1 & -73.7864 & -200.055  &  8.18278 \\ \hline
 2 &  49.9803 &    23.1201 & 1.62347 \\ \hline
 3 &  20.2636 &   -7.6633 &  0.187535 \\ \hline
 4 &   6.09059&    1.87212&  0.011658 \\ \hline
 5 &   1.42424 &  -0.374493& 0.00052486\\ \hline
 6 &   0.271056&  0.0622017  & 0.0000284767  \\ \hline
\end{tabular}
\end{center}}
\subsubsection*{Semilocal strings with $h_0=0.6$}
{\small\begin{center}
\textit{Seeley coefficients: Semilocal strings} \\
\begin{tabular}{|c|c|c|c|}
 \hline  & \multicolumn{3}{|c|}{$h_0=0.6$\, , \, $R=10^2$}  \\ \hline
 $n$ & ${\rm tr}\,c_n^I$ & ${\rm tr}\,c_n^O$ & $c_n^G$  \\ \hline
 1 &107.840 &  -113.330 &  13.3675 \\ \hline
 2 & 72.5374  &  41.052 &   1.73758  \\ \hline
 3 & 26.5220  & -12.8641&   0.135265 \\ \hline
 4 &  7.45753   & 3.13219&  0.00625589 \\ \hline
 5 &  1.67277 &  -0.615441& 0.000232254 \\ \hline
 6 &  0.309679 &  0.101414& 8.66613$\times 10^{-6}$ \\ \hline
\end{tabular}
\begin{tabular}{|c|c|c|c|}
 \hline  & \multicolumn{3}{|c|}{$h_0=0.6$ \, , \, $R=10^3$}  \\ \hline
 $n$ & ${\rm tr}\,c_n^I$ & ${\rm tr}\,c_n^O$ & $c_n^G$  \\ \hline
 1 &  132.751  & -189.607  &    14.4510 \\ \hline
 2 &   97.6535 &   65.4471    &  1.91781  \\ \hline
 3 &   34.6838 &  -20.9178 &     0.153289 \\ \hline
 4 &   9.48185 &    5.15141  &   0.0075433 \\ \hline
 5 &   2.0767 &    -1.01896 &    0.000303781 \\ \hline
 6 &   0.376957 &   0.168681 &   0.0000119181\\ \hline
\end{tabular}
\begin{tabular}{|c|c|c|c|}
 \hline  & \multicolumn{3}{|c|}{$h_0=0.6$ \, , \, $R=10^4$}  \\ \hline
 $n$ & ${\rm tr}\,c_n^I$ & ${\rm tr}\,c_n^O$ & $c_n^G$  \\ \hline
 1 & 157.661  & -265.885 &     9.98213 \\ \hline
 2 & 122.770  &   89.8409 &     1.92331    \\ \hline
 3 &  42.8451  &  -28.9709 &    0.21752 \\ \hline
 4 &  11.50061 &    7.17046 &    0.0137912 \\ \hline
 5 &   2.48061  &  -1.42244 &    0.000645635 \\ \hline
 6 &   0.44232 &    0.235944   &0.0000151754 \\ \hline
\end{tabular}
\begin{tabular}{|c|c|c|c|}
 \hline  & \multicolumn{3}{|c|}{$h_0=0.6$ \, , \, $R=10^5$}  \\ \hline
 $n$ & ${\rm tr}\,c_n^I$ & ${\rm tr}\,c_n^O$ & $c_n^G$  \\ \hline
 1 & 182.571 & -342.162    & 16.6184 \\ \hline
 2 & 147.886 &  114.235   &   2.27905 \\ \hline
 3 &  51.0065 & -37.0239   &  0.189413 \\ \hline
 4 &  13.5303 &  9.18952  &   0.0101236 \\ \hline
 5 &   2.88453 &-1.82592 &    0.000447104 \\ \hline
 6 &   0.511506&  0.303206 & 0.0000184335 \\ \hline
\end{tabular}

\begin{tabular}{|c|c|c|c|}
 \hline  & \multicolumn{3}{|c|}{$h_0=0.6$ \, , \, $R=10^6$}  \\ \hline
 $n$ & ${\rm tr}\,c_n^I$ & ${\rm tr}\,c_n^O$ & $c_n^G$  \\ \hline
 1 & 207.481 & -418.440 &  17.7021 \\ \hline
 2 & 173.003 &  138.629 &   2.45961\\ \hline
 3 &  59.1167 & -45.0769 &  0.207444 \\ \hline
 4 &  15.5545&   11.2086&   0.0114354 \\ \hline
 5 &   3.28844 & -2.2294&   0.000517096\\ \hline
 6 &   0.578786&  0.370467  & 0.0000217957  \\ \hline
\end{tabular}
\end{center}}
\subsubsection*{Semilocal strings with $h_0=0.9$}

{\small\begin{center}
\textit{Seeley coefficients: Semilocal strings} \\
\begin{tabular}{|c|c|c|c|}
 \hline  & \multicolumn{3}{|c|}{$h_0=0.9$\, , \, $R=10^2$}  \\ \hline
 $n$ & ${\rm tr}\,c_n^I$ & ${\rm tr}\,c_n^O$ & $c_n^G$  \\ \hline
 1 & 638.887 &  -601.992 &  17.3563 \\ \hline
 2 & 326.423  &  290.227 &   1.24111  \\ \hline
 3 & 109.814  & -94.6224&    0.0897853 \\ \hline
 4 &  27.9842 &  23.3075&    0.00588282 \\ \hline
 5 &  5.71851 &  -4.59264&   0.000325247 \\ \hline
 6 &  0.97486 &  0.755329&   0.0000152661 \\ \hline
\end{tabular}
\begin{tabular}{|c|c|c|c|}
 \hline  & \multicolumn{3}{|c|}{$h_0=0.9$ \, , \, $R=10^3$}  \\ \hline
 $n$ & ${\rm tr}\,c_n^I$ & ${\rm tr}\,c_n^O$ & $c_n^G$  \\ \hline
 1 &  1055.500  & -1030.530  &    25.0050 \\ \hline
 2 &   536.449 &    495.158    &   2.5130  \\ \hline
 3 &   178.399 &   -162.381 &      0.216979 \\ \hline
 4 &    45.0018 &    40.2880  &    0.014969 \\ \hline
 5 &     9.11549&    -7.798641 &   0.00083022 \\ \hline
 6 &     1.54073 &    1.32206 &   0.0000282526\\ \hline
\end{tabular}
\begin{tabular}{|c|c|c|c|}
 \hline  & \multicolumn{3}{|c|}{$h_0=0.9$ \, , \, $R=10^4$}  \\ \hline
 $n$ & ${\rm tr}\,c_n^I$ & ${\rm tr}\,c_n^O$ & $c_n^G$  \\ \hline
 1 & 1472.190  & -1459.110 &     32.6701 \\ \hline
 2 &  746.482  &   700.081 &      3.79044    \\ \hline
 3 &  246.919  &  -230.135 &      0.344728\\ \hline
 4 &   62.0179 &     57.2676 &    0.0240939 \\ \hline
 5 &   12.5121  &   -11.3800 &    0.00133716 \\ \hline
 6 &    2.10653 &     1.88677 &   0.0000612957 \\ \hline
\end{tabular}
\begin{tabular}{|c|c|c|c|}
 \hline  & \multicolumn{3}{|c|}{$h_0=0.9$ \, , \, $R=10^5$}  \\ \hline
 $n$ & ${\rm tr}\,c_n^I$ & ${\rm tr}\,c_n^O$ & $c_n^G$  \\ \hline
 1 & 1888.870 & -1887.670    & 40.3354 \\ \hline
 2 &  956.515 &   905.004   &   5.06804 \\ \hline
 3 &  315.440 &  -297.889   &   0.472483 \\ \hline
 4 &   79.0340 &   74.2472   &  0.0332125 \\ \hline
 5 &   15.2088 &  -14.7736 &    0.000184412 \\ \hline
 6 &    2.67232&    2.45247 &   0.0000843383 \\ \hline
\end{tabular}

\begin{tabular}{|c|c|c|c|}
 \hline  & \multicolumn{3}{|c|}{$h_0=0.9$ \, , \, $R=10^6$}  \\ \hline
 $n$ & ${\rm tr}\,c_n^I$ & ${\rm tr}\,c_n^O$ & $c_n^G$  \\ \hline
 1 & 2305.550 & -2316.230 &  48.007 \\ \hline
 2 & 1166.550 &  1109.930 &   6.34561\\ \hline
 3 &  383.961 &  -365.644 &   0.60024 \\ \hline
 4 &   96.05501&   91.2268&   0.0423323 \\ \hline
 5 &   19.3054&   -18.1672&   0.0423323\\ \hline
 6 &    3.23313&    3.01818  & 0.000107346  \\ \hline
\end{tabular}
\end{center}}

\section*{Acknowledgements} This work has been partially financed
by the Spanish Ministerio de Educacion y Ciencia and the Junta de
Castilla y Leon under grants: FIS2006-09417, and VA013C05.


\end{document}